\documentclass[11pt,draftcls,onecolumn]{article}
\usepackage{graphicx,xcolor}
\usepackage{epstopdf,bm,epsfig,amsthm,color}
\usepackage{bbm,amsmath,amsfonts,amssymb,empheq}
\usepackage[linesnumbered,ruled]{algorithm2e}
\usepackage{subcaption}
\usepackage{float}

\usepackage{geometry}
\def \ba {{\mathbf a}}

\def \bp {{\mathbf p}}
\def \bs {{\mathbf s}}

\begin{document}
	\title{Reinforcement Learning for Caching with Space-Time Popularity Dynamics}

	\author{
		Alireza Sadeghi, Georgios B. Giannakis, Gang Wang, and Fatemeh Sheikholeslami\thanks{A. Sadgehi,  G. B. Giannakis, and G. Wang are with the Digital Technology Center and the Department of Electrical and Computer Engineering, University of Minnesota, Minneapolis, MN 55455, USA. F. Sheikholeslami is with the Bosch Center for Artificial Intelligence, Pittsburgh, PA 15222, USA. E-mails: \{sadeghi, georgios, gangwang, sheik081\}@umn.edu.}
	}
	
\date{}

	\maketitle
	
	\allowdisplaybreaks


With the tremendous growth of data traffic over wired and wireless networks along with the increasing number of rich-media applications,  caching is envisioned to play a critical role in next-generation networks. To intelligently prefetch and store contents, a cache node should be able to learn what and when to cache. Considering the geographical and temporal content popularity dynamics, the limited available storage at cache nodes, as well as the interactive influence of caching decisions in networked caching settings, developing  effective caching policies is practically challenging. In response to these challenges, this chapter presents a versatile reinforcement learning based approach for near-optimal caching policy design, in both single-node and network caching settings under dynamic space-time popularities. The herein presented policies are complemented using a set of numerical tests, which showcase the merits of the presented approach relative to several standard caching policies.      

\section{Introduction}

The term cache was initially introduced in computer systems in $1970$s to describe a small capacity memory with practically fast access \cite{Paschosbook19, Paschos18, Bastug14}. This idea was later applied to the Internet, where content requests typically were routed to just a few central servers \cite{mag19wang, surveycache17}. Mimicking the computer caches, an Internet cache is to be deployed at the edge of the network, closer to end users, and thus to serve requests locally. Recently, the exponential growth of mobile video traffic due to the advent of smart phones, tablets, routers, and
a massive number of devices connected through the Internet of Things \cite{zanella2014internet,zhang2019mobile}, in conjunction with advances in Machine Learning and Artificial Intelligence has boosted caching to the front line of research in wired and wireless networks. In this regard, caching has been investigated from different perspectives. 

Several works have considered \textit{static} content popularities, which yields tractable caching models at the price of accuracy. For instance, a multi-armed bandit approach that accounts for the demand history and unknown popularities can be found in \cite{Gunduz}. Coded, convexified, and distributed extensions of this problem were studied in \cite{Sengupta}, context and trend-aware learning approaches in \cite{Schaar_2017}, \cite{Schaar_trend}, and coordinated-distributed extensions in \cite{Coordinated}. From a learning
perspective, the trade-off between the ``accuracy'' of learning a static popularity, and the corresponding learning ``speed'' was investigated \cite{Trade_off} and \cite{Learningbound}. 

Nonetheless, popularities in practice are \textit{dynamic}, meaning they fluctuate over a time horizon. For instance, half of the top $25$ requested Wikipedia articles change on a daily basis ~\cite{paschos2018,Wikidynamic}, which motivates well adaptive caching strategies that can account for popularity dynamics \cite{Cacherent, PSN1, PSN2, onlineCodedCaching, sadeghi19jsac, ITDeniz, Onlinerl19letter}. To approximate the evolution of popularities, a Poisson shot noise model was adopted in \cite{PSN1}, for which an age-based caching solution was proposed in \cite{PSN2}.  In addition, postulating a Markovian evolution for popularities, reinforcement learning based approaches were recently studied in \cite{rl195g, dist19rl, ddql20access, RL1, RL2}. Albeit reasonable for discrete states, these approaches cannot deal with large continuous state-action spaces. To cope with such spaces, deep reinforcement learning approaches have been considered for content caching in, e.g., \cite{DRL_19, DRL_AC}. 

The aforementioned works have focused on devising caching policies for a single entity. A more common setting
in next-generation networks however, involves a network of interconnected caching nodes. For example, today's content delivery
networks such as Akamai \cite{nygren2010akamai}, have tree network structures. On the other hand, it has been shown that optimizing a network of connected caches jointly can further improve performance \cite{Maddahali2014}. 

This chapter aspires to glean some of the recent advances in caching with space-time popularities, through a suite of reinforcement learning tools. The collection here is by no means exhaustive. Upon modeling the space-time popularity dynamics in a cellular network as a Markovian process, a reinforcement learning formulation of the caching task is laid out. After reviewing the classical reinforcement learning algorithms, $Q$-learning in particular, a scalable approach is derived. Next, a networked caching scenario is considered where a parent node is connected to several leaf nodes to serve end-user file requests. Inspired by the tree network structure in Akamai, a two-timescale formulation of the caching problem is developed. To model the interaction between caching decisions of parent and leaf nodes along with the space-time evolution of file requests, a scalable deep reinforcement learning approach based on hyper deep  $Q$-networks (DQNs) is developed.

\emph{Notation.} Lower- (upper-) case boldface letters denote column vectors (matrices), whose $(i,j)$-th entry is denoted by $[\,.\,]_{i,j}$. Calligraphic symbols are reserved for sets, while $\top$ stands for transposition. The operator $\mathbb E$ denotes the expectation, and vector $\bf 1$ represents the all one vector. 

\section{Single-node caching: Modeling and problem statement}\label{sec:model}
\begin{figure}[t]
	\centering
	\includegraphics[width=0.6\columnwidth]{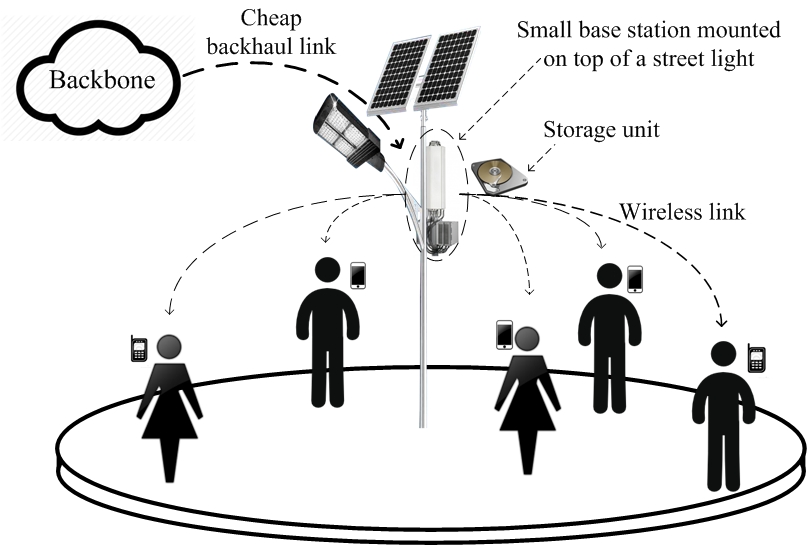}
	\caption{Local section of a HetNet.}
	\label{fig:sys_jstsp}
\end{figure}

\begin{figure}[t] 
	\centering
	{\includegraphics[width=.65\columnwidth]{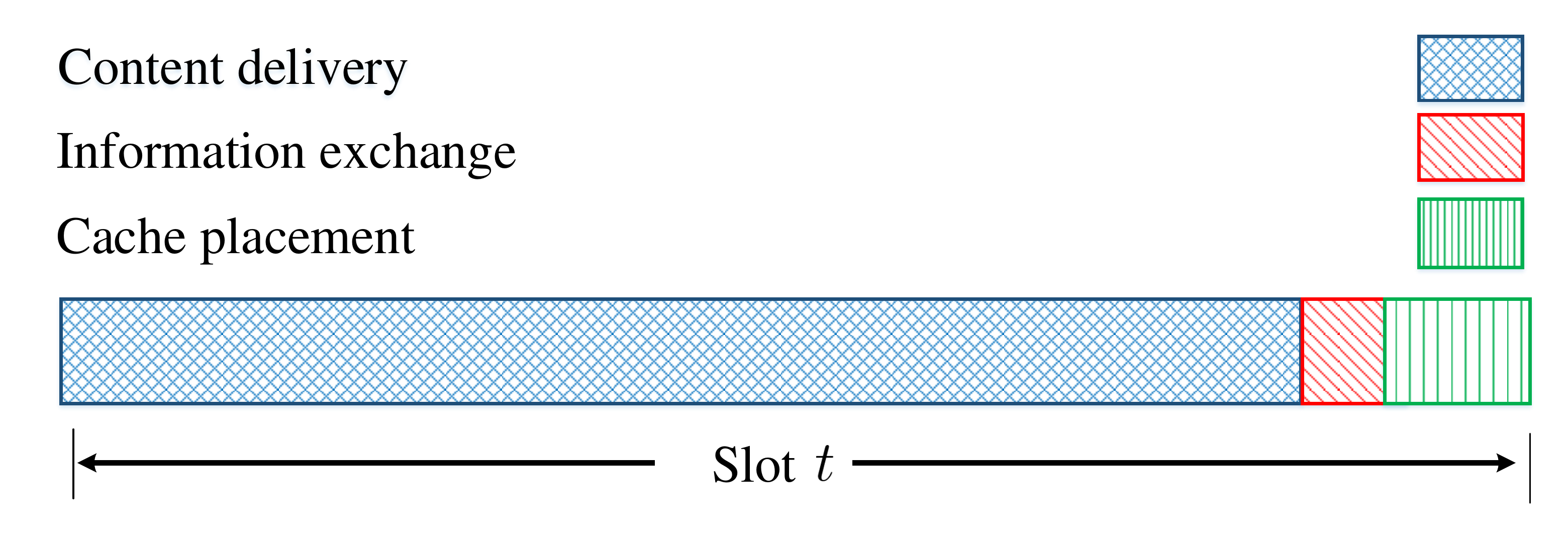}}
	\caption{The slot structure.} 
	\label{fig:slot} 	
\end{figure}

Consider a subsection of a cellular network, which is simply a single small base station (SB) connected to the backbone network through a low-bandwidth, high-delay backhaul link. The SB is equipped with $M$ units to store unit-size content (files); see Fig.~\ref{fig:sys_jstsp}. We suppose that caching is to be carried out in a slotted fashion over $t=1, \, 2, \,\ldots$, where, at the end of each slot, the cache control unit (CCU)-enabled SB selects ``intelligently'' $M$ files from a total of $F \gg M$ available ones, and prefetches them for possible reuse in the subsequent slot. The structure of a slot is depicted in Fig. \ref{fig:slot}. At the beginning of a time slot, the user file requests are revealed, and the so-called ``content delivery" phase takes place. The second phase, pertains to ``information exchange,'' where the SBs transmit their locally-observed popularity profiles to the network operator, and in return receive the estimated network-wide global popularity profile. Finally, ``cache placement'' is to happen where a selection of files are stored for the next time slot. These slots may not be of equal length, as the starting times may be set a priori, for example at 3~AM, 11~AM, or 4~PM, when the network load is relatively low; or, slot intervals may be dictated to the CCU by the network operator `on the fly.' Generally, a slot begins when the network is at an off-peak period, and its duration coincides with the peak traffic time when the pertinent costs of serving users are high.

Users request a subset of files from the set ${\cal F}:=\left\{1, 2,\ldots,F\right\}$ of available ones, during the content delivery of slot $t$. If a requested file is stored in the cache, it will be immediately served locally, without incurring any cost. Conversely, if it is not available in the cache, then the SB must fetch it from the cloud through a congested backhaul link, thus incurring a certain cost, due to possible electricity price surges, processing cost, or the sizable delay resulting in low QoE and user dissatisfaction. The goal is to enable the CCU to intelligently select the cache contents, so that costly services from the cloud can be avoided as often as possible. To this aim, let ${\bf a} (t) \in \mathcal{A}$ denote the $F \times 1$ binary \emph{caching action vector} at slot $t$, with $\mathcal{A}:=\{\mathbf{a}| \mathbf{a} \in \{0,  1\}^F, \mathbf{a}^\top \mathbf{1}=M\}$ being the set of all feasible actions; where having $[{{\bf a}(t)}]_f  = 1$ is tantamount to storing the file $f$ for the duration of slot $t$; and  $[{\bf a}(t)]_f  = 0$ otherwise.

During the content delivery phase, the CCU computes an $F \times 1$-vector of the \emph{local popularity profile} ${\bf p}_L (t)$ per slot $t$, whose $f$-th entry captures the expected local demand for file $f$, defined as
\begin{align}
\nonumber
\bigg[\mathbf{p}_{\text{L}} (t)\bigg]_f :=\dfrac{\text{ Number of local requests for } f \; {\text {at slot} } \;t }{\text{Number of all local requests at slot}\; t}.
\end{align}
Similarly, the network operator estimates an $F \times 1$ \emph{global popularity profile} vector ${\bf p}_G (t)$, and transmits it to~all~CCUs. Having the estimated local and global file popularities by the end of the information exchange phase of slot $t$, we define the overall system state as
\begin{align}
{\bf s} (t):= \left[{\bf p}^{\top}_G (t),{\bf p}^{\top}_L (t),{\bf a}^{\top}(t)\right]^{\top}.
\label{eq.1}
\end{align}
While at slot $t-1$, our \emph{objective} is to leverage historical observations of states $\left\{{\bf s} (\tau)\right\}_{\tau = 0}^{t-1}$, and corresponding costs, to learn an optimal action for the next slot; that is, ${\bf a}^{\ast}(t)$. The ensuing sections formulate this objective from a reinforcement learning vantage point. 

\subsection{Costs and caching strategies}
\label{Regorously_Problem_formulation}

\begin{figure}[t] 
	\centering
	{\includegraphics[width=.6\columnwidth]{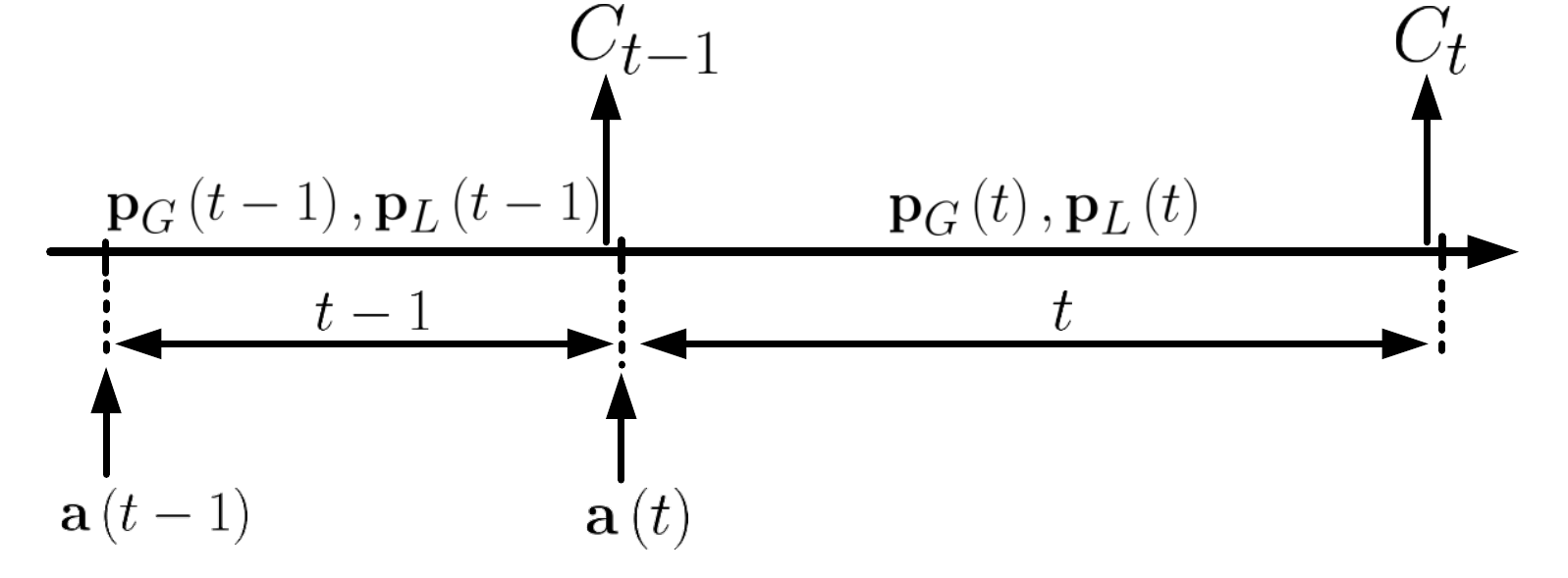}}
	\caption{A schematic depicting the evolution of key quantities across time slots. Duration of slots can be unequal.} 
	\label{fig:sys2} 	
\end{figure}

Serving file requests is costly, where the overall cost comprises the superposition of three terms.
\begin{subequations}	
The first term $c_{1,t}$ corresponds to the cost of refreshing the cache contents locally. In its  general form, $c_{1,t} (\cdot)$ is a function of the upcoming action $\mathbf{a}(t)$, and available contents at the cache according to current caching action $\mathbf{a}(t-1)$, where the subscript $t$ captures the possibility of incurring a time-varying cost for refreshing the cache. A reasonable choice of $c_{1,t}(\cdot)$ is
\begin{align}
c_{1,t}(\mathbf{a}(t),\mathbf{a}(t-1)) := \lambda_{1,t} \mathbf{a}^\top(t) \left[\mathbf{1}- \mathbf{a}(t-1)\right] \;.
\label{subcost1}
\end{align}
The second cost $c_{2,t}(\mathbf{s}(t))$ is incurred during the operational phase of slot $t$ to satisfy local user file requests. A prudent choice for this cost must: i)~penalize requests for files already cached much less than requests for files not stored; and, ii) be a non-decreasing function of popularities $[\mathbf{p}_L]_f$. For simplicity, we choose 
\begin{align}
c_{2,t}(\mathbf{s}(t)):= \lambda_{2,t} \left[\mathbf{1}-\mathbf{a}(t)\right]^\top \mathbf{p}_L(t)
\label{subcost2}
\end{align}
which solely penalizes the non-cached files in a descending order of their local {popularities}.

The third type of  cost captures the  ``mismatch'' between caching action $\mathbf{a}(t)$, and the global popularity profile $\mathbf{p}_{\text{G}}(t)$. Indeed, it is  reasonable to consider the global popularity of files as a surrogate of what the local profiles must look like in the near future; thus, keeping the caching action close to $\mathbf{p}_{\text{G}}(t)$ may reduce future possible costs. Note that having a relatively small number of local requests may just give a rough estimate of local demands, while the global popularity profile can serve as side information in tracking the evolution of content popularities across the network. 
To account for this issue, we introduce the third type of cost as
\begin{align}
c_{3,t}(\mathbf{s}(t)):= \lambda_{3,t} \left[\mathbf{1}-\mathbf{a}(t)\right]^\top \mathbf{p}_G(t)
\label{subcost3}
\end{align}
penalizing the files not cached according to the global popularity profile  ${\bf p}_G (\cdot)$ provided by the  network operator, thus promoting adaptation of caching policies close to global demand~trends.  
	
\end{subequations}
Upon taking action $\mathbf{a}(t)$, the \emph{aggregate cost conditioned} on the popularity vectors revealed, can be expressed as follows
\begin{align}\label{Overall_Cost}
 C_t \Big({ {\bf s} (t-1), {\bf a} (t) \Big| {\mathbf{p}_{\text{G}}(t)},{\mathbf{p}_{\text{L}}(t)}} \Big) & :=   c_{1,t}\left({\bf a} (t), {\bf a} (t-1)\right) + c_{2,t}\left({\bf s} (t)\right) + c_{3,t}(\mathbf{s}(t)) \\ \nonumber & = \lambda_{1,t} \mathbf{a}^\top(t) (\mathbf{1}- \mathbf{a}(t-1)) +\lambda_{2,t} (\mathbf{1}-\mathbf{a}(t))^\top \mathbf{p}_L(t)  \\& \quad+ \lambda_{3,t} (\mathbf{1}-\mathbf{a}(t))^\top \mathbf{p}_G(t)\nonumber
\end{align} 
where weights $\lambda_{1,t}$, $\lambda_{2,t}$, and $\lambda_{3,t}$ control the relative significance of the corresponding terms. As asserted earlier, the cache-refreshing cost is considered to be less than that of fetching, which justifies $\lambda_{1,t} \ll \lambda_{2,t}$. In addition, setting $\lambda_{3,t} \ll \lambda_{2,t}$ is of interest when  the local popularity profiles are of acceptable accuracy, or, if tracking local popularities is of higher importance. In particular, setting $\lambda_{3,t} = 0$ corresponds to the special case where the caching cost is  decoupled from the global popularity profile evolution. On the other hand, setting $\lambda_{2,t} \ll \lambda_{3,t}$  is desirable in networks where globally popular files are of high significance, for instance when
users have high mobility and may change SBs rapidly, or, when a few local requests prevent the SB from estimating accurately the local popularity profiles. Figure~\ref{fig:sys2} depicts the evolution of popularities and action vectors along with the aggregate conditional costs across slots. 

\subsection{Markovian popularity profile evolution}
\label{Dynamic evolution of user demands}
\begin{figure}[t]
	\centering
	{\includegraphics[width=.7\columnwidth]{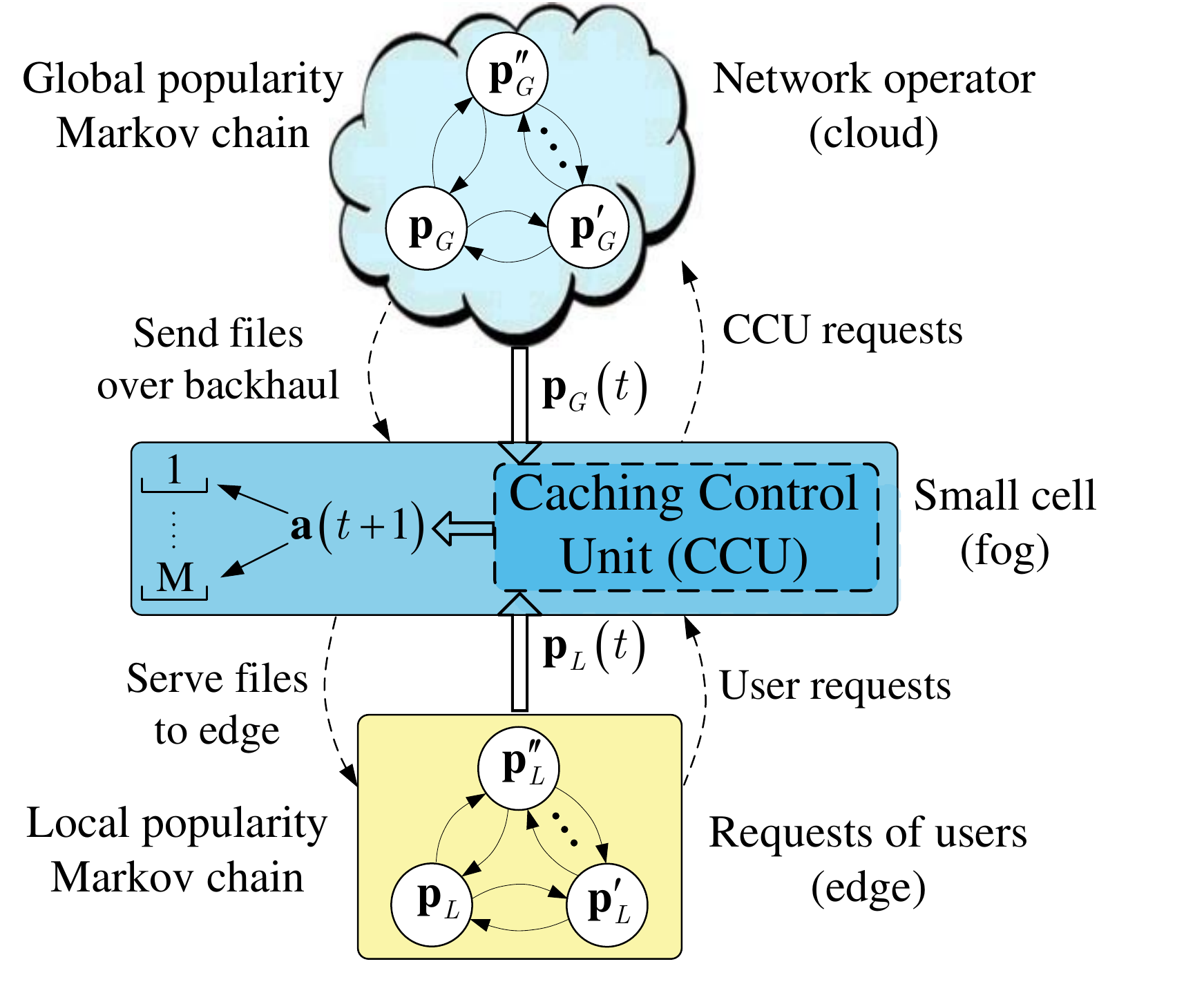}}
	\caption{Schematic of network structure with required communications between an SB with the network operator.} 
	\label{fig:sys3} 
\end{figure}

We consider Markovian evolution for user requests (popularities) both globally and locally as depicted in Fig.~\ref{fig:sys3}. Global popularity profiles are assumed to be generated through an underlying Markov chain with $|{\cal P}_G|$ states collected in the set 
${\cal P}_G :=\left\{ {\bf p}_{G}^{1}, \ldots, {\bf p}_{G}^{|{\cal P}_G|} \right\}$; and likewise, for the local popularity profiles having states
${\cal P}_L:=\left\{ {\bf p}_{L}^{1}, \ldots, {\bf p}_{L}^{|{\cal P}_L|} \right\}$. Although ${\cal P}_G$ and ${\cal P}_L$ are known, the underlying transition probabilities of the two Markov chains are unknown in practice. 

Given ${\cal P}_G$ and ${\cal P}_L$ as well as feasible caching decisions from the set $\cal A$, the overall set of states in the network is 
\begin{align}
\nonumber {\cal S} := \left\{{\bf s}\big| {\bf s}= [{\bf p}^{\top}_G, {\bf p}^{\top}_L, {\bf a}^{\top}]^{\top}, {\bf p}_G \in {\cal P}_G  \textrm{ , } {\bf p}_L \in {\cal P}_L, {\bf a} \in {\cal A} \right\}.
\end{align} 
In the proposed reinforcement learning based caching, the underlying transition probabilities for global and local popularity profiles are  \textit{unknown}. Therefore, the learner should find the optimal policy by interactively by making sequential decisions, and observing the corresponding costs. The following section formulates the optimal caching problem, and provides an efficient solver to handle the ``curse of dimensionality'' typically emerging with reinforcement learning  \cite{Sutton1998reinforcement}.

\subsection{Reinforcement learning formulation}	

Due to the random nature of user requests, the cost \eqref{Overall_Cost} is random with mean	
\begin{align}\label{mean_Cost}
{\overline C}_t \left( {\bf s} (t-1), {\bf a} (t) \right)  
 :&=\mathbb{E}_{{\mathbf{p}_{\text{G}}(t)},{\mathbf{p}_{\text{L}}(t)}} \Big[ C_t \left( { {\bf s} (t-1), {\bf a} (t) {\Big |} {\mathbf{p}_{\text{G}}(t)},{\mathbf{p}_{\text{L}}(t)}} \right) \Big] \\
&  = \lambda_1 \mathbf{a}^\top(t) \left[\mathbf{1}- \mathbf{a}(t-1)\right] +\lambda_2 \mathbb{E} \left[ (\mathbf{1}-\mathbf{a}(t))^\top \mathbf{p}_L(t)\right] \nonumber \\  & \quad + \lambda_3 \mathbb{E} \left[(\mathbf{1}-\mathbf{a}(t))^\top \mathbf{p}_G(t)\right] \nonumber
\end{align}
where the expectation is taken with respect to (wrt)  ${\bf p}_L (t)$ and ${\bf p}_G (t)$, while the weights are $\lambda_{1,t} = \lambda_1$, $\lambda_{2,t} = \lambda_2$, and $\lambda_{3,t} = \lambda_3$. Let $\pi: \mathcal{S} \rightarrow \mathcal{A}$ denote the policy function, which maps state $\mathbf{s} \in  \mathcal{S}$ to the action set, where for a given  $\pi(\cdot)$, at the state ${\bf s}(t)$, caching is carried out via $\mathbf{a}(t+1)=\pi(\mathbf{s}(t))$  dictating what files to store for next slot $(t+1)$. The caching performance is measured through the so-termed state value function 
\begin{align}\label{eq_Value_function}
V_{\pi} \left({\bf s} (t)\right) :=   \lim\limits_{T \rightarrow \infty} \mathbb{E}  \left[\sum\limits_{\tau = t}^{T} \gamma^{\tau-t} {\overline C}\left({\bf s} \left[\tau\right], {\pi} \left(\mathbf{s}\left[\tau\right]\right) \right)\Big|{\bf s} (t)\right] 
\end{align}
which is the expected discounted cost incurred over an infinite time horizon, with future terms discounted by a factor  $\gamma \in \left[0,1\right)$. Furthermore, tuning $\gamma$ trades off current for future costs, and it also accounts for modeling uncertainties or dynamics. 

The objective here is to seek an optimal policy $\pi^*$ such that the expected discounted cost is minimized for any initial state ${\bf s}$  
\begin{align}
\label{eq_Opt_policy}
{\pi^{\ast}} = \arg \min \limits_{\pi \in \Pi} V_{\pi} \left( {\bf s} \right), \quad \forall {\bf s} \in {\cal S}
\end{align} where $\Pi$ is the set of all feasible policies. The  optimization problem in  \eqref{eq_Opt_policy} is a sequential decision-making problem, which can be solved using the Bellman equations, as elaborated in next section.

\subsection{Optimality conditions}
\label{Bellman}
The Bellman equations provide necessary conditions for optimality of a policy in a sequential decision making problem. Let $[{\bm P}^a]_{{\bf s}{\bf s}'}$ denote the transition probability of going from the current state ${\bf s}$ to the next one ${\bf s}'$ while having taken action ${\bf a}$; that is, 
\begin{align}
[{\bm P}^{\bf a}]_{{\bf s}{\bf s}'} := {\rm{Pr}} \Big \{{{\bf s}(t) = {\bf s}'}  {\Big |} {\bf s}(t-1) = {\bf s}, \pi(\mathbf{s}(t-1)) = \mathbf{a}\Big \}.
\end{align}  
Bellman equations express the state value function by \eqref{eq_Value_function} in a recursive fashion as follows \cite[pg. 47]{Sutton1998reinforcement}
\begin{align}\label{eq_Value_function_recurring}
V_{\pi} \left({\bf s} \right) = { \overline{C}\left({\bf s}, \pi(\mathbf{s})  \right)  }+ \gamma  \sum_{{\bf s}' \in \mathcal{S}} [{\bm P}^{\pi({\bf s})}]_{{\bf s}{\bf s}'}   V_{\pi} \left({\bf s}' \right)\;, \forall {\bf s,s}'
\end{align}
which amounts to the superposition of $\overline C$ plus a discounted version of future state value functions under a given policy $\pi$. Specifically, ${ \overline{C}}$ in \eqref{mean_Cost} can be written as 

\begin{align}
\label{cbardef}
{ \overline{C}\left({\bf s}, \pi(\mathbf{s})  \right)  } = \sum_{\bs'  := [\bp_G',\bp_L',\ba']\in \mathcal{S}}  [{\bm P}^{\pi({\bf s})}]_{{\bf s}{\bf s}'}  C\Big({\bf s}, \pi(\mathbf{s}) \Big| \bp_G',  \bp_L'   \Big)  
\end{align}
where $ C\big({\bf s}, \pi(\mathbf{s}) \big| \bp_G',  \bp_L'   \big)$ is defined in \eqref{Overall_Cost}.
Given $[{\bm P}^{\bf a}]_{{\bf s}{\bf s}'}$ $\forall {\bf s,s}'$, one can readily obtain $\left\{V_{\pi} ({\bs}), \forall {\bs} \right\}$ by solving \eqref{eq_Value_function_recurring}, and eventually the optimal policy $\pi^{\ast}$ in \eqref{pi*} using the so-termed policy iteration algorithm \cite[pg. 79]{Sutton1998reinforcement}. In practice however, ${\bm P}^{\bf a}$ is unknown. This motivates the use of adaptive dynamic programming (ADP) and model-free reinforcement learning algorithms such as $Q$-learning. The ADP estimates $[{\bm P}^{\bf a}]_{{\bf s}{\bf s}'}$ for all ${\bf s, s}' \in \mathcal{S}$, and ${\bf a} \in \mathcal{A}$, as iterations proceed \cite[pg. 834]{AIModernapproach}. Unfortunately, ADP algorithms are often very slow and impractical, as they must estimate $|\mathcal{S} |^2 \times |\mathcal{A}|$ probabilities. In contrast, the  $Q$-learning algorithm to be elaborated next finds the optimal $\pi^{*}$ as well as $V_{\pi}({\bf s})$, while circumventing the need for estimating $[{\bm P}^a]_{{\bf s}{\bf s}'}, \forall {\bf s,s}'$; see e.g., \cite[pg.~140]{Sutton1998reinforcement}.

To describe how this algorithm works in our context, we introduce the so-termed state-action value function for a given policy $\pi$ \cite[pg. 62]{Sutton1998reinforcement}
\begin{align}
\label{Q_pi}
Q_{\pi} \left({\bf s} , {{\ba}'}\right) :=    { \overline{C}\left({\bf s},  {\ba'}  \right)}  +  \gamma  \sum_{{\bf s}' \in \mathcal{S}} [{\bm P}^{{\ba'}}]_{{\bf s}{\bf s}'}   V_{\pi} \left({\bf s}' \right) 
\end{align}
which is commonly referred to as the ``Q-function''. The  $Q$-factor  $Q_{\pi}({\bf s},{\bf a})$ basically captures the expected current cost of taking a particular action ${\bf a}$ when the system is in state $\bf s$, followed by the discounted value of the future states, provided that the future actions are taken by following the policy~$\pi$. The next section uses the $Q$-function to find the optimal policy.

 \subsection{Optimal caching via  $Q$-learning}
 \label{Q-learning}
  $Q$-learning is a reinforcement learning scheme to infer the optimal policy $\pi^{\ast}$, by estimating the optimal state-action value function  $Q^*(\mathbf{s,a}')  := Q_{\pi^{\ast}}(\mathbf{s,a}') ,\; \forall {\bf s},{\bf a}'$ `on the fly.' Upon obtaining $Q^*(\mathbf{s,a}')$, the optimal policy can be easily extracted  \cite[pg. 67]{Sutton1998reinforcement} as 
 \begin{align}\label{pi*}
 \pi^*({\bf s}) = \arg\min_{{\boldsymbol \alpha}}~Q^{*}({\bf s}, {\boldsymbol \alpha}), \quad \forall {\bf s} \in {\cal S}.
 \end{align}
 Furthermore, the  $Q$-function and  $V(\cdot)$ under $\pi^{\ast}$ are related 
 \begin{align}\label{v*}
 V^{*}({\bf s}):= V_{\pi^*}({\bf s}) =\min_{{\boldsymbol \alpha}} Q^{*}({\bf s}, {\boldsymbol \alpha})
 \end{align}
 which in turn yields 
 \begin{align}
 \label{eq_Q_function2}
 Q^* \left({\bf s} , {\bf a}'\right) =  \overline{C}\left({\bf s}, {\bf a}'  \right) +  \gamma  \sum_{{\bf s}' \in \mathcal{S}} [{\bm P}^{\bf a}]_{{\bf s}{\bf s}'}  \min_{{\boldsymbol \alpha}\in {\cal A}} Q^* \left({\bf s'} , {\boldsymbol \alpha}\right).
 \end{align} 
 The well-known $Q$-learning algorithm relies on \eqref{eq_Q_function2} to approximate $Q^*(\mathbf{s,a}')$. This algorithm is tabulated in Algorithm \ref{alg:Q_learning}. 
 In this algorithm, the agent updates its estimated $\hat{Q}(\mathbf{s}(t-1),\mathbf{a}(t))$ as  $C\big( {{\bf s} (t-1), {\bf a} (t)} \big |  {\mathbf{p}_{\text{G}}(t)},{\mathbf{p}_{\text{L}}(t)} \big)$ is observed. That is, given ${\bf s}(t-1)$,  $Q$-learning takes action ${\mathbf{a}}(t)$, and  upon observing ${\mathbf{s}}(t)$, it incurs cost and forms the instantaneous error 
 \begin{align}\label{eq:error}
 \varepsilon \left({\bf s}(t-1), {\bf a}(t)\right) := \frac{1}{2} \Big( C\left({\bf s}(t-1),{\bf a}(t)\right) + \gamma \min \limits_{{\boldsymbol \alpha}}^{} {\widehat Q} \left({\bf s}(t), {\boldsymbol \alpha}\right)  - {\widehat Q} \left({\bf s}(t-1),{\bf a}(t)\right) \Big)^2 
 \end{align} 
and correspondingly updates the  $Q$-function through a stochastic gradient descent step to minimize error. As a result, one can easily write the update rule in short as   
 \begin{align}
 \hat{Q}_t\left({\bf s}(t-1),{\bf a}(t)\right) & =  (1-\beta_t) \hat{Q}_{t-1}\left({\bf s}(t-1),{\bf a}(t)\right)   \nonumber  \\ 
 \nonumber
 & \quad + \beta_t  \Big[C\left( {{\bf s} (t-1), {\bf a} (t)} {\Big |}  {\mathbf{p}_{\text{G}}(t)},{\mathbf{p}_{\text{L}}(t)} \right) + \gamma \min_{{\boldsymbol \alpha}} \hat{Q}_{t-1}\left({\bf s}(t),{{\boldsymbol \alpha}}\right) \Big] \nonumber 
 \end{align}
while keeping the rest of the entries in $\hat{Q}_t(\cdot,\cdot)$ unchanged.

\begin{algorithm}[t]
	\caption{Caching via  $Q$-learning at CCU}
	\label{alg:Q_learning}
		{\bf Initialize}  $\mathbf{s}(0)$ randomly and $\hat{Q}_0(\mathbf{s,a}) = 0 \; \forall \mathbf{s,a}$ \\
		\For  {$t = 1,2,... $}{{
		Take action ${\bf a}(t) $ chosen probabilistically by \[{\bf a}(t)  = \left\{
			\begin{array}{ll}
			\arg \min \limits_{\bf a} {\hat Q}_{t-1}\left({\bf s} (t-1),{\bf a}\right) & \textrm{w.p. }\;\; 1-\epsilon_t \\
			\textrm{random } \mathbf{a} \in \mathcal{A}  & \textrm{w.p.} \;\; \; \epsilon_t
			\end{array}
			\right. \]} 
		 {${\bf p}_L (t)$ and ${\bf p}_G (t)$ are revealed based on user requests} \\
		{Set ${\bf s} (t) = \left[{\bf p}_G^{\top} (t), {\bf p}_L^{\top} (t) , {\bf a}(t)^{\top}\right]^{\top}$} \\
		{Incur cost $C\Big({ {\bf s} (t-1), {\bf a} (t) {\Big |} {\mathbf{p}_{\text{G}}(t)},{\mathbf{p}_{\text{L}}(t)}}\Big)$}
		\\ Update \hfill
		\begin{align}
		\hat{Q}_t\!\left({\bf s}(t-1),{\bf a}(t)\right) &= (1-\beta_t) \hat{Q}_{t-1}\left({\bf s}(t-1),{\bf a}(t)\right)   \nonumber \\ 
		\nonumber
		 \quad  & +\beta_t  \Big[C\!\left( {{\bf s} (t-1), {\bf a} (t)} {\big |}  {\mathbf{p}_{\text{G}}(t)},{\mathbf{p}_{\text{L}}(t)} \right) + \gamma \min_{{\boldsymbol \alpha}} \hat{Q}_{t-1}\left({\bf s}(t),{{\boldsymbol \alpha}}\right) \Big] \nonumber 
		\end{align}
	}
\end{algorithm} 

Regarding convergence of the  $Q$-learning algorithm, a necessary condition ensuring ${\hat Q}_{t}\left(\cdot,\cdot\right) \rightarrow {Q}^{*}\left(\cdot,\cdot\right)$, is  that  all state-action pairs must be continuously updated \cite{WatkinsQ}. Under this and the usual stochastic approximation conditions that will be specified later, ${\hat Q}_t \left(\cdot,\cdot\right)$ converges to ${Q}^{*} \left(\cdot,\cdot\right)$ with probability~$1$; see~e.g., \cite{tsitsiklis},
for a detailed description. Finite-time error bounds of the $Q$-learning algorithm with function approximation can be found in \cite{wang2019multistep},  \cite{aistats2020sun}.  
 To guarantee visiting all state-action pairs, various exploration-exploitation algorithms have been proposed. In this section, we have adopted an $\epsilon$-greedy action selection algorithm. 
During initial iterations, or when the CCU observes a considerable shift in content popularities, setting $\epsilon_t$ high promotes exploration in order to learn the underlying dynamics. On the other hand, in stationary settings and once ``enough'' observations are made, small values of $\epsilon_t$ are desirable as they enable agent actions to approach the optimal policy.


The main limitation of the  $Q$-learning algorithm is its slow convergence, which is due to the independent updates of the $Q$-values. Fortunately, $Q$-values are related, and leveraging these relationships leads to multiple updates per observation, hence accelerating convergence. In the ensuing section, the structure of the problem at hand is exploited  to develop a linear function approximation of the $Q$-function, which in turn offers scalability. 

\subsection{Scalable caching}
\label{Linear Function approximation}
Despite the simplicity of the updates as well as the optimality guarantees of the  $Q$-learning algorithm, it is not scalable due to the possibly large state-action spaces present in real networks. For instance, the  $Q$-table is of size $|\mathcal{P}_G| |\mathcal{P}_L| |\mathcal{A}|^2$, where  $|\mathcal{A}|={F \choose M} $ encompasses all possible selections of $M$ out of $F$ files. Hence, the  $Q$-table size grows prohibitively with $F$, rendering the convergence of the table entries, as well as the policy iterates unacceptably slow. Furthermore, action selection in $\min_{{\boldsymbol \alpha} \in {\cal A}} Q({\bf s},{\bf a})$ entails an expensive exhaustive search over the entire feasible action set $\mathcal{A}$ whose cardinality can be huge.  Function approximation schemes are appealing as they can endow  $Q$-learning with scalability to handle real-world problems \cite{linear18rl, geramifard2013tutorial,mahadevan2009learning,haj2019deep, meanfield18}.  
In our problem, a delicate linear approximation for $Q(\mathbf{s},\mathbf{a})$ is inspired by the additive form of the instantaneous costs in \eqref{Overall_Cost}. Specifically, we propose to approximate $Q(\mathbf{s},\mathbf{a}')$ as follows   
\begin{align}
\label{approximation}
Q(\mathbf{s},\mathbf{a}')\simeq Q_G(\mathbf{s},\mathbf{a}')+Q_L(\mathbf{s},\mathbf{a}')+Q_R(\mathbf{s},\mathbf{a}')
\end{align}
where $Q_G$, $Q_L$, and $Q_R$ correspond to the global and local popularity mismatches, and the cache-refreshing cost, respectively.  Recall that the state vector $\mathbf{s}$ consists of three subvectors as $\mathbf{s} := [{\mathbf p}^{\top}_G, {\mathbf p}^{\top}_L,\mathbf{a}^{\top}]^{\top}$. Corresponding to the global popularity subvector, we model the first term of the approximation in \eqref{approximation} as
\begin{align}
\label{Q_G_est_1}
Q_G(\mathbf{s},\mathbf{a}'):= \sum_{i=1}^{|\mathcal{P}_G|}\sum_{f=1}^F \theta^{G}_{i,f} \mathbbm{1}_{\left\{\mathbf{p}_G={\bf p}^{i}_{G}\right\}} \mathbbm{1}_{\left\{[{\ba}']_f=0\right\}}
\end{align}	
where the sums are over possible global popularity profiles as well as  files, and where the indicator function ${\mathbbm{1}}_{\left\{ \cdot \right\}}$ is $1$ if its argument holds true; and $0$ otherwise; while $\theta^{G}_{i,f}$ captures the average ``overall'' cost if the system is in global state ${\bf p}^{i}_G$, and the CCU decides not to cache the $f$th content. By defining the  $|{\cal P}_G| \times |\cal F|$ matrix with $(i,f)$-th entry $\left[{\boldsymbol \Theta}^G\right]_{i,f} := \theta^{G}_{i,f}$, one can rewrite \eqref{Q_G_est_1} as 
\begin{align}
\label{Matrix_G}
Q_G(\mathbf{s},\mathbf{a}')= \boldsymbol{\delta}_G^{\top}({\bp_G}) \boldsymbol{\Theta}^G (\mathbf{1}-{\bf a}')
\end{align}  
where 
\begin{align} \nonumber
{\boldsymbol \delta}_G ({\bf p}_G) := \left[\delta({\bf p}_G-{\bf p}^{1}_{G}), \ldots, \delta({\bf p}_G-{\bf p}^{|{\cal P}_G|}_{G})\right]^{\top}\:.
\end{align} 
Similarly, we can approximate the second summand  \eqref{approximation} as 
\begin{align}
\nonumber
Q_L(\mathbf{s},\mathbf{a}')&:= \sum_{i=1}^{|\mathcal{P}_L|}\sum_{f=1}^F \theta^{L}_{i,f} \mathbbm{1}_{\left\{\mathbf{p}_L={\bf p}^{i}_{L}\right\}} \mathbbm{1}_{\left\{[{\ba}']_f=0\right\}} \\ \label{Q_L_est_2} &   = \boldsymbol{\delta}^\top_{L} ({{\bp}_L}) \boldsymbol{\Theta}^L (\mathbf{1}-{\bf a}')
\end{align}
where $\left[{\boldsymbol \Theta}^L\right]_{i,f} := \theta^{L}_{i,f}$, and  
\begin{align} \nonumber
{\boldsymbol \delta}_L ({\bf p}_L) := \left[\delta({\bf p}_L-{\bf p}^{1}_{L}), \ldots, \delta({\bf p}_L-{\bf p}^{|{\cal P}_L|}_{L})\right]^{\top}
\end{align}
with $\theta^{L}_{i,f}$ modeling the average overall cost for not caching file $f$ when the local popularity is in state ${\bf p}^i_L$.

Finally, we model the third summand in \eqref{approximation} corresponding to the cache refreshing cost, as follows
\begin{align} \label{eq.Qr}
Q_R(\mathbf{s},\mathbf{a}'):&= \sum_{f=1}^F \theta^{R} \pmb{1}_{\left\{[{\ba}']_f=1\right\}} \pmb{1}_{\left\{[\ba]_f=0\right\}} \\ & = \theta^{R} {\bf a}'^{\top} \left(1-{\bf a}\right)  \nonumber \\ &= \theta^{R} \left[ {\bf a}'^{\top} \left(1-{\bf a}\right) +{\bf a}^{\top} {\bf 1} - {\bf a}'^{\top} {\bf 1}\right] \nonumber
\\ &= \theta^{R} {\bf a}^{\top} (\mathbf{1}-{\bf a}') \nonumber 
\end{align}
where $\theta^R$ is the long-time averaged cache-refreshing cost per content. The constraint {$\mathbf{a}^\top \mathbf{1}= \mathbf{a}'^\top \mathbf{1} = M$},
is to factor out the term $\bf{1-a}'$, which will become useful shortly.

Upon collectively denoting all parameters by $\Lambda := \{\boldsymbol{\Theta}^G,\boldsymbol{\Theta}^L, \theta^a\}$, the  $Q$-function is readily approximated as (cf. \eqref{approximation}) 
\begin{align}
\label{eq.app}
{\widehat Q}_{\Lambda}(\bs,{\ba}'):= \underbrace{\Big( \boldsymbol{\delta}_G^{\top}({\bp_G}) \boldsymbol{\Theta}^G +  \boldsymbol{\delta}_L^{\top}({\bp_L}) \boldsymbol{\Theta}^L    + \theta^{R} {\bf a}^{\top} \Big)}_{\boldsymbol \psi(\bf s):=} (\mathbf{1}-{{\ba}'}).
\end{align}
The original task of learning $|{\cal P}_G| |{\cal P}_L| |{\cal A}|^2$ parameters in Algorithm \ref{alg:Q_learning} is now reduced to that of learning $\Lambda$ containing $\left( \left|{\cal P}_G\right| + \left|{\cal P}_L\right| \right) \left|{\cal F}\right|  +1$ parameters.

\subsection{Learning $\Lambda$}
Given the current parameter estimates  $\{\widehat{\boldsymbol{\Theta}}_{t-1}^G,\widehat{\boldsymbol{\Theta}}_{t-1}^L, \hat{\theta}_{t-1}^R\}$ at the end of the information exchange phase of slot $t$, the so-called temporal difference error is given by 
\begin{align} \nonumber 
\widehat{e} \left({\bf s}(t-1), {\bf a}(t)\right)  :=   C\left({\bf s}(t-1),{\bf a}(t)\right) + \gamma \min \limits_{{\bf a}'}^{} {\widehat Q}_{{\Lambda_{t-1}}} \left({\bf s}(t), {\bf a}'\right) - {\widehat Q}_{{\Lambda_{t-1}}} \left({\bf s}(t-1),{\bf a}(t)\right) . 
\end{align} 
Using the definition 
\begin{align}
\label{eq:errorfncapp}
\widehat\varepsilon \left({\bf s}(t-1), {\bf a}(t)\right) := \dfrac{1}{2} \Big(\widehat{e}\left({\bf s}(t-1), {\bf a}(t)\right)\Big)^2,
\end{align}
the parameter update rules are obtained following stochastic gradient descent to minimize the loss in \eqref{eq:errorfncapp} as follows (cf. \cite[p. 847]{AIModernapproach})
\begin{align}
\label{UpdatethetaG}
\hat{\boldsymbol{\Theta}}_t^G & = \hat{\boldsymbol{\Theta}}_{t-1}^G - \alpha_G \nabla_{{\boldsymbol{\Theta}}^G} \widehat\varepsilon \left({\bf s}(t-1), {\bf a}(t)\right) 
\\ & = \hat{\boldsymbol{\Theta}}_{t-1}^G + \alpha_G \;{\widehat{e}  \left({\bf s}(t-1), {\bf a}(t)\right) } \; \;  \nabla_{{\boldsymbol \Theta}^{G}} {\widehat Q}_{{\Lambda_{t-1}}} ({\bf s}(t-1),{\bf a}(t))\nonumber \\  &
= \hat{\boldsymbol{\Theta}}_{t-1}^G + \alpha_G \;{\widehat{e}  \left({\bf s}(t-1), {\bf a}(t)\right) } \; \;   \boldsymbol{\delta }_G ({\bp_G(t-1)}) (\mathbf{1}-\ba(t))^\top \nonumber
\end{align} 
and
\begin{align}
\label{UpdatethetaL}
 \hat{\boldsymbol{\Theta}}_t^L &= \hat{\boldsymbol{\Theta}}_{t-1}^L - \alpha_L \nabla_{{\boldsymbol{\Theta}}^L} \widehat\varepsilon \left({\bf s}(t-1), {\bf a}(t)\right) 
\\ & = \hat{\boldsymbol{\Theta}}_{t-1}^L + \alpha_L \;{\widehat{e}  \left({\bf s}(t-1), {\bf a}(t)\right) } \; \;  \nabla_{{\boldsymbol \Theta}^{L}} {\widehat Q}_{{\Lambda_{t-1}}} ({\bf s}(t-1),{\bf a}(t))\nonumber \\  
& = \hat{\boldsymbol{\Theta}}_{t-1}^L + \alpha_L \;{\widehat{e}  \left({\bf s}(t-1), {\bf a}(t)\right) } \; \;   \boldsymbol{\delta }_L ({\bp_L(t-1)}) (\mathbf{1}-\ba(t))^\top \nonumber
\end{align} 
along with 
\begin{align}
\label{UpdatethetaR}
& \hat{{\theta}}_t^R = \hat{{\theta}}_{t-1}^R - \alpha_R \nabla_{{{\theta}}^R} \widehat\varepsilon \left({\bf s}(t-1), {\bf a}(t)\right)
\\ &= \hat{{\theta}}_{t-1}^R + \alpha_R \;{\widehat{e}  \left({\bf s}(t-1), {\bf a}(t)\right) } \; \;  \nabla_{{\theta}^R} {\widehat Q}_{\Lambda_{t-1}} ({\bf s}(t-1),{\bf a}(t))\nonumber \\  
&= \hat{{\theta}}_{t-1}^R + \alpha_R \;{\widehat{e}  \left({\bf s}(t-1), {\bf a}(t)\right) } \; \;  {\bf a}^{\top}(t-1) (\mathbf{1}-{\bf a}(t)) \nonumber.
\end{align} 
The pseudocode for this scalable approximation of the  $Q$-learning scheme is tabulated as Algorithm \ref{alg:qlearnapprox}. 

\begin{algorithm}[t]
	\caption{Scalable  $Q$-learning}
	\label{alg:qlearnapprox}
		{\bf Initialize}  $\mathbf{s}(0)$ randomly, ${\widehat{\boldsymbol \Theta}_0^G} = {\bf 0}$, ${\widehat{\boldsymbol \Theta}_0^L} = {\bf 0}$, ${\hat \theta}_0^R = 0$, and thus $\widehat{\boldsymbol{\psi}}(\bf s) = {\bf 0}$ \\
		\For  {$t=1,2,...$}
		{{Take action ${\bf a}(t) $ chosen probabilistically by \[{\bf a}(t)  = \left\{
			\begin{array}{ll}
			\text {$M$ best files via ${\widehat{\boldsymbol \psi} \left({\bf s}(t-1)\right)}$} & \textrm{w.p. }\;\; 1-\epsilon_t \\
			\textrm{random } \mathbf{a} \in \mathcal{A}  & \textrm{w.p.} \;\; \; \epsilon_t
			\end{array}
			\right. \]  \hspace{0.4 cm} where 	 ${\boldsymbol{ \hat{\psi}} ({\bf s})} := \boldsymbol{\delta}_G^{\top}({\bp_G}) \boldsymbol{\widehat{\Theta}}^G +  \boldsymbol{\delta}_L^{\top}({\bp_L}) \boldsymbol{\widehat{\Theta}}^L    + \widehat{\theta}^{R} {\bf a}^{\top}$}
		\\
		${\bf p}_G (t)$ and ${\bf p}_L (t)$ are revealed based on user requests\\
		Set \quad  \hspace{0.9 cm} ${\bf s} (t) = \left[{\bf p}_G^{\top} (t), {\bf p}_L^{\top} (t) , {\bf a}(t)^{\top}\right]^{\top}$
		\\
		Incur cost \quad $C\Big({ {\bf s} (t-1), {\bf a} (t) {\Big |} {\mathbf{p}_{\text{G}}(t)},{\mathbf{p}_{\text{L}}(t)}}\Big)$ 
		\\
		Find \hspace{1 cm} $\widehat \varepsilon \left({\bf s}(t-1),{\bf a}(t)\right)$
		\\
		Update \quad \hspace{0.3cm} ${\widehat{\boldsymbol{\Theta}}}_t^G$, ${\widehat{\boldsymbol{\Theta}}}_t^L$ and ${\hat \theta}_t^R$ based on \eqref{UpdatethetaG}-\eqref{UpdatethetaR}
		}
\end{algorithm}              
The advantage of this scalable $Q$-learning is threefold.
\begin{itemize}
	\item The large state-action  space in the  $Q$-learning algorithm is handled by reducing the number of parameters from $|{\cal P}_G| |{\cal P}_L| |{\cal A}|^2$ to  $\left( \left|{\cal P}_G\right| + \left|{\cal P}_L\right| \right) \left|{\cal F}\right| +1$.
	\item In contrast to single-entry updates in the exact  $Q$-learning Algorithm \ref{alg:Q_learning} , $(F-M)$ entries in  $\widehat{\boldsymbol\Theta}^G$ and $\widehat{\boldsymbol\Theta}^L$ as well as $\theta^R$, are updated per observation using \eqref{UpdatethetaG}-\eqref{UpdatethetaR}, which leads to a much faster convergence.
	\item The exhaustive search in $ \min \limits_{{\bf a} \in {\cal A}} Q \left({\bf s},{\bf a}\right)$ required in exploitation; and also in the error evaluation \eqref{eq:error}, is circumvented. Specifically, it holds that (cf.~\eqref{eq.app}) \begin{align}
	\label{a}
	\min \limits_{{\bf a}' \in {\cal A}} Q({\bf s},{\bf a}') \approx \min \limits_{{\bf a}' \in {\cal A}} {\boldsymbol \psi}^{\top} ({\bf s}) \left({\bf 1- a}' \right) = \max  \limits_{{\bf a}'\in {\cal A}} {\boldsymbol \psi}^{\top}({\bf s}) \, \, {\bf a}'
	\end{align}
	where ${\boldsymbol{ \psi} ({\bf s})} := \boldsymbol{\delta}_G^{\top}({\bp_G}) \boldsymbol{\Theta}^G +  \boldsymbol{\delta}_L^{\top}({\bp_L}) \boldsymbol{\Theta}^L    + \theta^{R} {\bf a}^{\top}.$
	The solution of \eqref{a} is readily given by $[{\bf a}]_{\nu_i} = 1$  for $i=1, \ldots, M$, and $[{\bf a}]_{\nu_i} = 0$ for $i>M$, where $\left[{{\boldsymbol{ \psi} ({\bf s})}}\right]_{\nu_F} \le \cdots \le  \left[{{\boldsymbol{ \psi} ({\bf s})}}\right]_{\nu_1} $ are sorted entries of ${\boldsymbol{ \psi} ({\bf s})}$.  
\end{itemize} 

\subsection{Numerical tests}
This section tests the performance of the proposed $Q$-learning algorithm and its scalable approximation. To compare the proposed algorithms with the optimal caching policy, which is the best policy computed under \textit{known} transition probabilities for global and local popularity Markov chains, we first simulated a small network with $F = 10$ contents, and caching capacity $M = 2$ at the local SB. Global popularity profile is modeled by a two-state Markov chain with states ${\bf p}^{1}_G$ and
${\bf p}^{2}_G$, that are drawn from Zipf distributions having parameters $\eta_1^G=1$ and $\eta_2^G=1.5$, respectively \cite{breslau1999web}. That is, for state $i \in \left\{1,2\right\}$, the $F$ contents are assigned a random ordering of popularities, and then sorted accordingly in a descending order. Given this ordering and the Zipf distribution parameter $\eta_i^G$, the popularity of the $f$-th  content is set to  
\begin{align}
\nonumber
\big[\mathbf{p}_{\text{G}}^{i}\big]_f = \frac{1}{f^{{\eta^G_i}} \sum \limits_{l=1}^{F}   1 \mathbin{/} l^{\eta_i^G}}, \;\quad \;\;i=1,2 \end{align} 
where the summation normalizes the components to follow a valid probability mass function, while $\eta_i^G \geq 0$ controls the skewness of popularities. 

Furthermore, state transition probabilities  of the global popularity Markov chain were drawn randomly. 
Similarly, local popularities were modeled by a two-state Markov chain, with states $\mathbf{p}_L^{1}$ and  $\mathbf{p}_L^{2}$, whose entries were drawn from Zipf distributions with parameters $\eta^{L}_1 = 0.7$ and $\eta^{L}_2 = 2.5$, respectively. The transition probabilities of the local popularity Markov chain were generated randomly.    
\begin{figure}[t]
	\centering
	\includegraphics[width=.6\columnwidth]{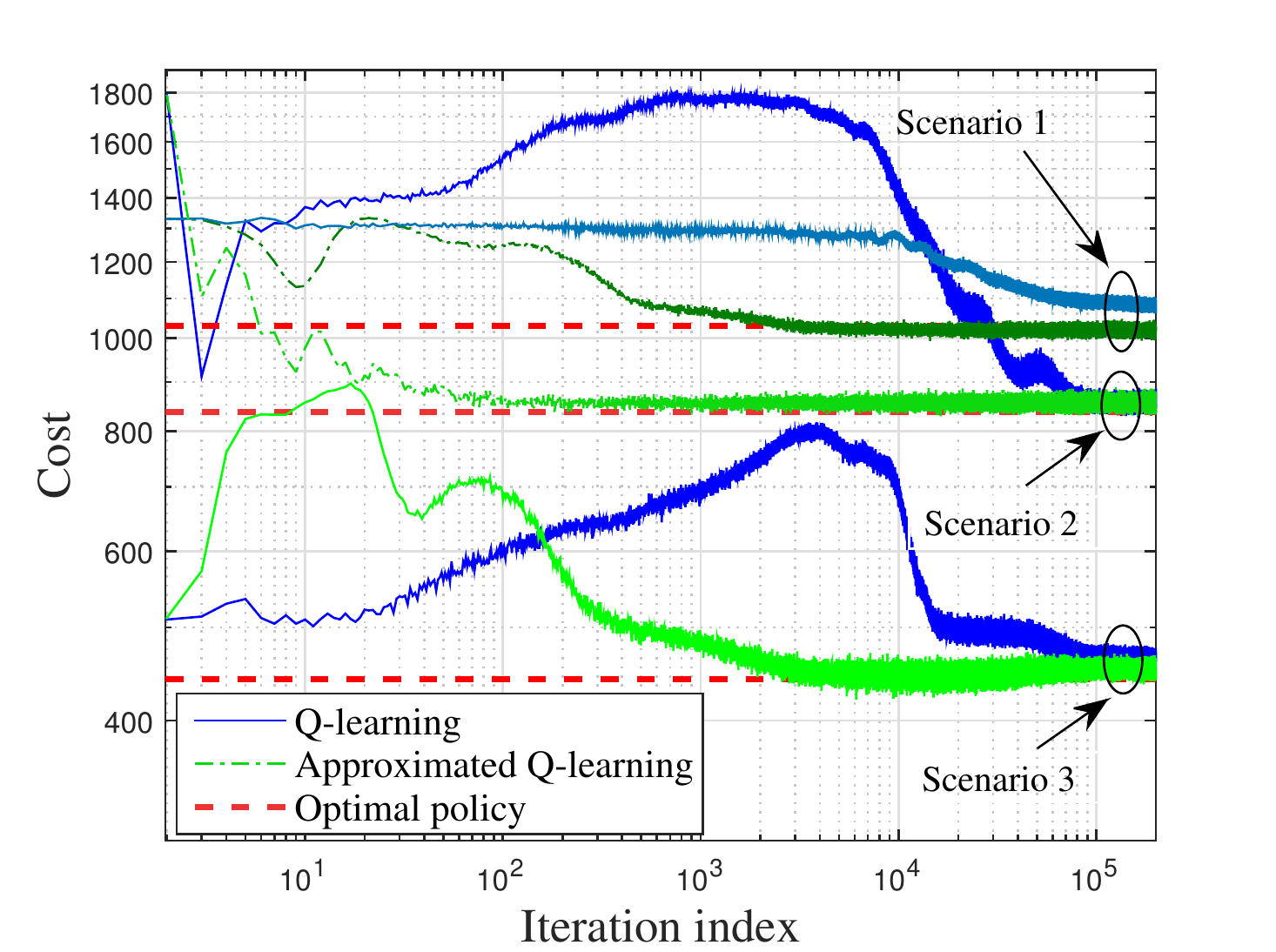}
	\caption{Performance of the proposed algorithms.}
	\label{Cost}
\end{figure}

Caching performance was assessed under three cost-parameter settings: (s1)   $\lambda_1 = 10, \lambda_2 = 600, \lambda_3 = 1000$; (s2) $\lambda_1 = 600, \lambda_2 = 10, \lambda_3 = 1000$, and (s3) $\lambda_1 = 10, \lambda_2 = 10, \lambda_3 = 1000$. In all numerical tests the optimal caching policy was found by utilizing the policy iteration algorithm with known transition probabilities. In addition,  $Q$-learning in Algorithm  \ref{alg:Q_learning} and its scalable approximation in Algorithm  \ref{alg:qlearnapprox} were run with $\beta_t = 0.8$, $\alpha_{G} = \alpha_{L} = \alpha_{R} = 0.005$, and $\epsilon_t = 0.05$.

Figure~\ref{Cost} depicts the observed cost versus iteration (time) index averaged over 1,000 realizations. It is seen that the caching cost via  $Q$-learning, as well as through its scalable approximation converges to that of the  optimal policy. As anticipated,  even for this small-size network, with $|{\cal P}_G| = |{\cal P}_L| = 2$ and $|{\cal A}| = 45$, the  $Q$-learning algorithm converges slowly to the optimal policy, especially under s1, while its scalable approximation exhibits faster convergence. The reason for the slower convergence under (s1) is that the corresponding cost parameters of local and global popularity mismatches are set high, and thus convergence of the  $Q$-learning algorithm as well as that of the caching policy essentially relies on learning both the global and local popularity Markov chains. In contrast, under (s2), $\lambda_2$ corresponding to the local popularity mismatch is low; hence, the impact of local popularity Markov chain on the optimal policy is reduced, giving rise to a simpler policy, thereby a faster convergence.

Having demonstrated the accuracy and efficiency of the proposed algorithms, we next simulated a larger network with $F = 1,000$ available files, and a cache capacity of $M = 10$, offering a total of ${{1000}\choose{10} }\simeq 2 \times 10^{23}$ feasible caching actions. In addition, we set the local and global popularity Markov chains to have $|{\cal P}_L| = 40$  and $|{\cal P}_G| = 50$ states, for which the underlying state transition probabilities were drawn randomly, and Zipf parameters were drawn uniformly over the interval $(2,4)$. 

\begin{figure}[t]
	\centering
	\includegraphics[width=.6\columnwidth]{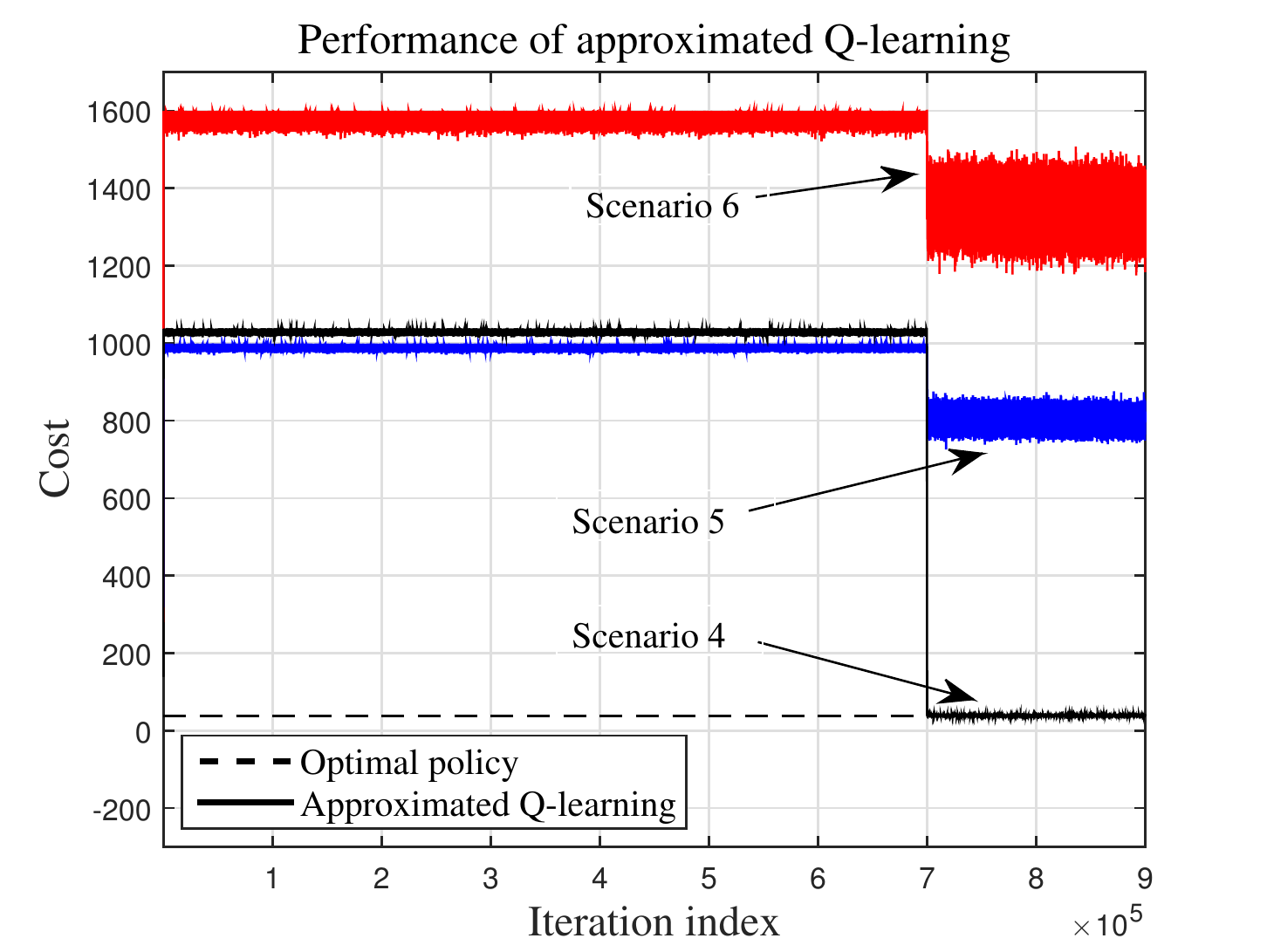}
	\caption{Performance  in large state-action space scenaria.}
	\label{appr}
\end{figure}

Figure~\ref{appr} plots the performance of Algorithm  \ref{alg:qlearnapprox} under (s4) $\lambda_1 = 100$, $\lambda_2 =20$, $\lambda_3 =20$, (s5) $\lambda_1 = 0$, $\lambda_2 = 0$, $\lambda_3 = 1,000$, and (s6)~$\lambda_1 = 0$, $\lambda_2 = 1,000$, $\lambda_3 = 600$. The exploration-exploitation parameter was set to  $\epsilon_t=1 $ for $t=1,2,\ldots,7 \times 10^{5}$, in order to greedily explore the entire state-action space in initial iterations, and $\epsilon_t = 1 \mathbin{/} ({\textrm{iteration index}})$ for $t>7 \times 10^{5}$.  

Finding the optimal policy in (s5) and (s6) requires a prohibitively sizable memory as well as incurs an extremely high computational complexity. It is thus impractical for this network. However, having large cache-refreshing costs with $\lambda_1 \gg \lambda_2,\lambda_3$ in (s4) forces the optimal caching policy to freeze its cache contents, making the optimal caching policy predictable in this setting. Despite the very limited storage capacity of $10\mathbin{/}1,000=0.01$ of available files, reinforcement learning-enabled caching offers considerable cost reduction, while the proposed approximated  $Q$-learning endows the approach with scalability and light-weight updates. 

The proposed Algorithms \ref{alg:Q_learning} and \ref{alg:qlearnapprox} deal with only a single caching entity. However, in more common settings, a cache node resides within a network of interconnected caches. For example, most of today's content delivery networks such as Akamai have a tree network structure \cite{nygren2010akamai}. Therefore, the caching policy design for a network of interconnected caches has become common practice in recent contributions; see, e.g. \cite{DRL_19, collaborative2012}. To design practical caching policies for such networks, one is faced with the following challenges, including i) topology-adaptive caching: caching decisions of a node in a network of caches, influence decisions of all other nodes, and thus a desired caching policy must adapt to the network topology as well as policies of the neighboring nodes; ii) complex dynamics: content popularities are random, and  exhibit unknown space-time, heterogeneous, and often non-stationary dynamics over the entire network; and, iii) large and continuous state-space: because of the sheer size of available content, caching nodes, and possible realizations of requests, the decision space is huge. These considerations prompted us to address the network caching problem from a reinforcement learning perspective. 

\section{Network caching with space-time popularity dynamics}

\begin{figure}
	\centering
	{\includegraphics[width=0.4\textwidth]{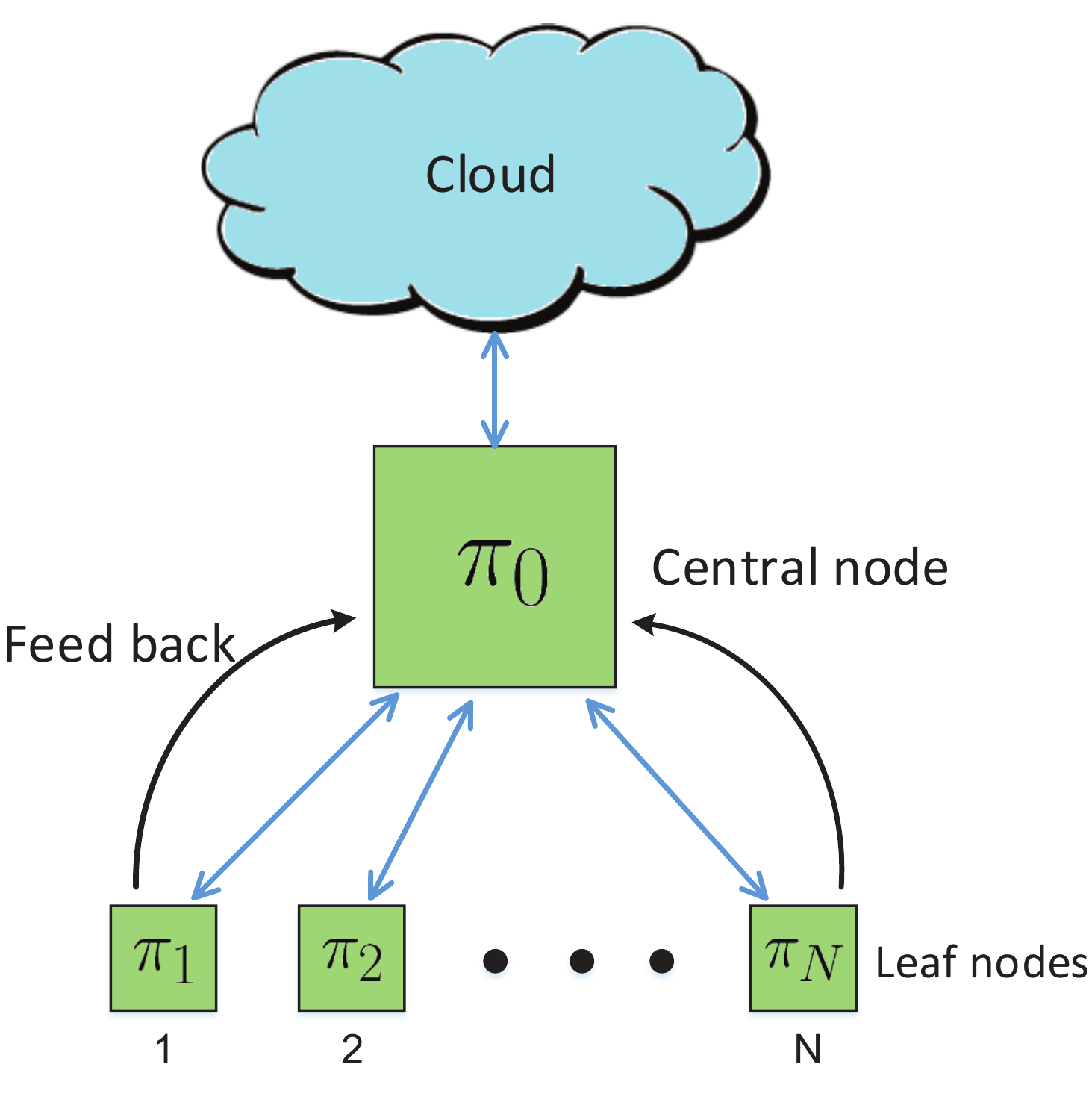}}
	{\includegraphics[width=0.4\textwidth]{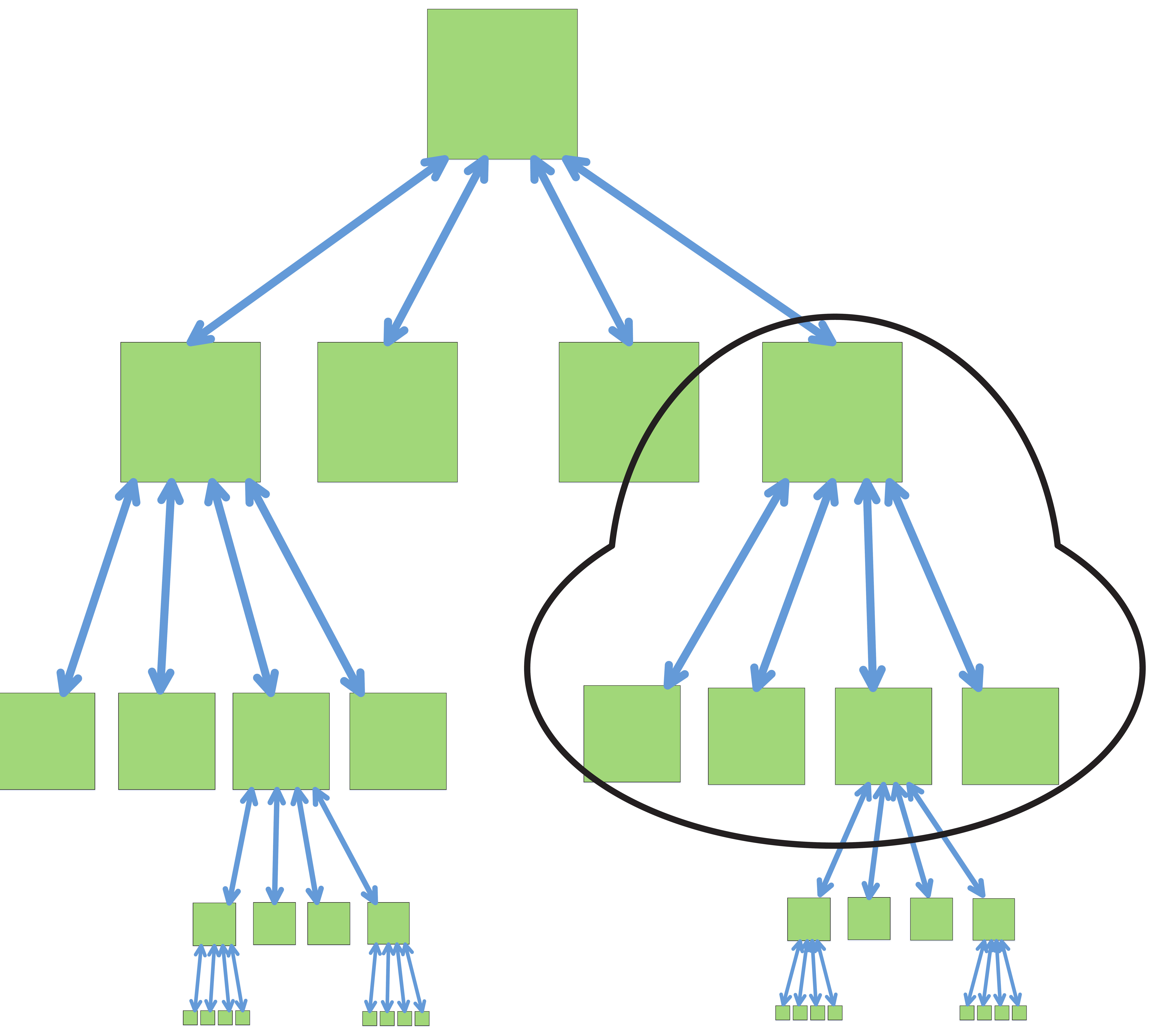}}
	\caption{The considered network of caching nodes (left); and,
		a hierarchical tree network of caches (right).}
	\label{fig:model_tccn}
\end{figure}

Consider a two-level network caching, where a parent node is connected to multiple leaf nodes to serve end-user file requests. Indeed, this two-level network constitutes a building block of the popular tree hierarchical cache networks in e.g., \cite{nygren2010akamai}; see also Fig.~\ref{fig:model_tccn} for an illustration.  To capture the complicated interactions between caching decisions of parent and leaf nodes along with the space-time evolution of file requests, we develop a scalable deep reinforcement learning approach based on a hyper deep $Q$-network (DQN) implementation developed in our prior works \cite{sadeghi19drl, qiu19drl}. Indeed, deep reinforcement learning has demonstrated state-of-the-art performance in diverse domains, including e.g., video games \cite{minh2015}, data centers \cite{smartect,ijcai2020}, smart grid \cite{qiu19drl}, and software-defined networking \cite{marvel,jsac2020}.
The objective here is to endow the caching policy of the parent node with the capability of adapting itself to local policies of leaf nodes and space-time evolution of file requests. 

\subsection{Modeling and problem statement}\label{sec:model_tccn}

Consider the two-level network of interconnected caching nodes in Fig. \ref{fig:model_tccn} (left), where a parent node is connected several leaf nodes indexed by $n\in\mathcal{N}:=\{1,\ldots,N\}$. The parent node is also connected to the cloud through a congested back-haul link. One could consider this network as part of a larger hierarchical caching system, where the parent node is connected to a higher level caching node instead of the cloud, as depicted in Fig.~\ref{fig:model_tccn} (right); see also \cite{ramadan2019framework, babaie2019cache}. 

All nodes in this network store files to serve file requests. Every leaf node serves its locally connected end users, by providing their requested files. If a requested content is  available locally at a leaf node, the content will be served immediately at no cost. If it is not locally available due to limited caching capacity, the content will be fetched from its parent node, at a certain cost. Similarly, if the file is available at the parent node, it will be served to the leaf at no cost; otherwise, the file must be fetched from the cloud at a higher cost.

Every leaf node follows its local caching policy to store locally popular files, while the parent node should store globally popular files that typically are not stored by the leaf nodes. Leaf nodes are closer to end users; therefore, they frequently receive user file requests that exhibit rapid temporal evolution at a \emph{fast} timescale. In contrast, the parent node observes an aggregate of requests over a large number of users that are not served by the $N$ leaf nodes, which naturally exhibit smaller fluctuations and thus evolve at a \emph{slow} timescale. 

\begin{figure}
	\centering
	\includegraphics[width =0.8 \textwidth]{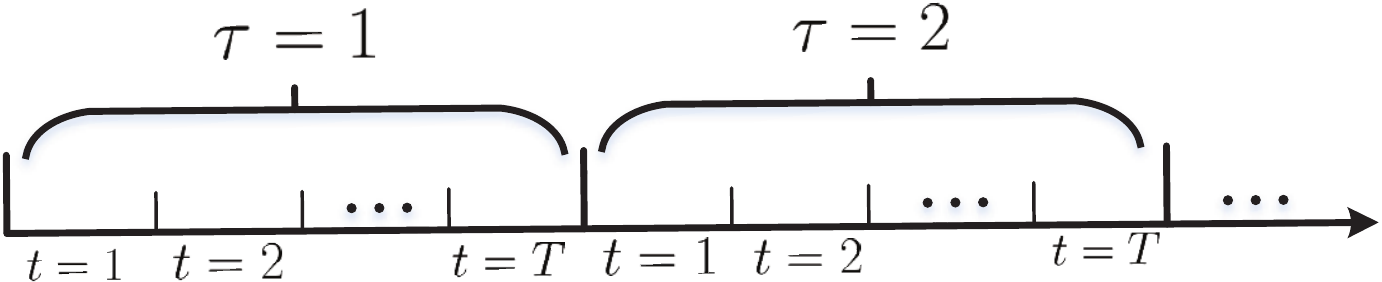}
	\caption{Slow and fast time slots.}
	\label{fig:Timescales}
	\vspace{-5pt}
\end{figure}

Building on this observation, we consider a two-timescale approach to managing such a tree network of caching nodes. To that end, let $\tau=1,2,\ldots$ denote the slow time intervals, each of which is further divided into $T$ fast time slots indexed by $t=1,\ldots,T$; see Fig.~\ref{fig:Timescales} for a depiction. We assume that the network state remains unchanged during each fast time slot $t$, but can change from $t$ to $t+1$. Suppose that the $F$ files in the cloud are collected in the set ${\mathcal F} = \{1,  \ldots, F\}$. At the beginning of each slot $t$, every leaf node $n$ selects a subset of files in $\mathcal{F}$ to prefetch and store for possible use in this slot. To determine which files to store, every leaf node relies on a local caching policy function $\pi_n$, to take action ${\pmb a}_n ( t+1, \tau  ) = \pi_n (\pmb s_n (t, \tau))$ at the beginning of slot $t+1$, based on its \emph{state} vector ${\pmb s}_n$ at the end of slot $t$. We define the state vector $\pmb s_n (t, \tau)\! :=\! \pmb r_n(t,\tau) := [r_n^1(t,\tau) \cdots r_n^{F}(t,\tau)]^\top$ to collect the number of requests received at leaf node $n$ for individual files over the duration of slot $t$ on interval $\tau$. Likewise, to serve file requests that have not been served by leaf nodes, the parent node takes action $\pmb a_0 ( \tau )$ to store files at the beginning of every interval $\tau$, according to a certain policy $\pi_0$. 

\subsection{Two-timescale problem formulation}\label{sec:two}
File transmission over links consumes resources, including e.g., energy, time, and bandwidth. Among possible choices, we consider the following cost for node $n\in\mathcal{N}$, at slot $t+1$ of interval $\tau$ to serve a request
\begin{multline}
\hspace{-0.3 cm} \pmb c_{n}(\pi_{n} (\pmb s_n ( t,\tau ) ), \pmb r_{n} (t+1,\tau), \pmb a_0 (	\tau ) )  \! :=  \! \pmb r_n (t+1, \tau) \odot  ( {\bf 1}\! - \pmb a_{0} (\tau))  \\  \!\odot \! ( {\bf 1} \!- \pmb a_{n} (t\!+\!1,\tau) )\! + \pmb r_n (t\!+\!1, \tau) \odot ( {\bf 1} \!- \pmb a_n(t\!+\!1,\tau) )
\label{eq:nodalcost}
\end{multline}
where $\pmb c_{n} (\cdot):=[c_n^1(\cdot)~\cdots~c_n^F(\cdot)]^{\top}$ concatenates the cost for serving individual files per node $n$; symbol $\odot$ denotes entry-wise vector multiplication; entries of $\pmb a_0$ and $\pmb a_n$ are either $1$ (cache, hence no need to fetch), or, $0$ (no-cache, hence fetch); and $\bf 1$ stands for the all-one vector. Specifically, the second term in \eqref{eq:nodalcost} is the cost of the leaf node fetching files for end users, while the first term is that of the parent fetching from the cloud. 

We again model file requests as Markov processes with unknown transition probabilities. During any interval $\tau$, a reasonable caching policy for leaf node $n\in\mathcal{N}$ minimizes the expected cumulative cost; that is, 
\begin{align} 
\pi^{ \ast}_{n,\tau} \!\!:= \underset{\pi_n \in \Pi_n}{\arg \min} ~ {\mathbb{E}}\Big[ \!\sum_{t=1}^{T}\! {\bf 1}^{\!\top} \!\pmb c_{n} (\pi_{n} (\pmb s_n ( t,\tau ) ), \pmb r_{n} (t\!+\!1,\tau), \pmb a_0 (\tau ))\Big] 
\label{eq:leaf_node}
\end{align} 
where $\Pi_n$ represents the feasible policies for node~$n$. Although solving \eqref{eq:leaf_node} is in general challenging, efficient near-optimal solutions have been introduced~\cite{RL1,RL2,multiarm2014}.  The remainder will thus focus only on designing the caching policy $\pi_0$ for the parent node.

To have a caching policy that accounts for both leaf node policies and file popularity dynamics, the parent collects relevant local information from all leaf nodes. At the end of interval $\tau$, each leaf node $n\in\mathcal{N}$ first obtains the time-average state vector ${\bar {\pmb s}_n} (\tau):= (1/T) \sum_{t=1}^{T}  \pmb s_{n} ( t , \tau)$; and at the same time forms and forwards the per-node vector ${\bar {\pmb s}_{n}} ( \tau ) \odot ({\bf 1} - \pi_n ( {\bar {\pmb s}_{n}} ( \tau )) )$ to its parent node. This vector has average number of file requests received during interval $\tau$, and zero entries if $\pi_n$ stores the corresponding files. Using the latter, the parent node forms its weighted state vector
\begin{align}
\label{eq:netstate}
\pmb s_0 (\tau):= \sum_{n = 1}^{N} w_n {\bar {\pmb s}_{n}} ( \tau ) \odot ({\bf 1} - \pi_n \!\left( {\bar {\pmb s}_{n}} (\tau) \right) )
\end{align}
with the weights $\{w_n\ge 0\}$. Similarly, having received at the end of interval $\tau +1$ the slot-averaged costs
\begin{multline}
{\bar {\pmb c}_n}( \pi_0 (\pmb s_0 ( \tau))  ;  \{ \pmb a_n ( t, \tau +1), \pmb r_n ( t , \tau +1) \}_{t=1}^T )\\ 
= \frac{1}{T} \sum_{t = 1}^{T} \pmb c_{n}(\pmb a_{n} (t,\tau+1), \pmb r_{n} \!\left(t,\tau+1 \right), \pmb a_0 (	\tau +1) )\label{eq:cost_pernode}
\end{multline}
from all leaf nodes, the parent node computes the cost 
\begin{align} \label{eq:c0}
\pmb c_0 ( {\pmb s}_0 ( \tau), \pi_0 (\pmb s_0 ( \tau)) )  = \sum_{n=1}^{N} w_n  {\bar {\pmb c}_n}( \pmb \pi_0 ( \pmb s_0 (\tau) )  ; \{ \pmb a_n ( t, \tau +1), \pmb r_n ( t , \tau +1) \}_{t=1}^T ) ). 
\end{align}
Having observed $\left\{\pmb s_0( \tau' ), \pmb a_0 ( \tau'), \pmb c_0 ( {\pmb s}_0 ( \tau'-1), \pmb a_0 (\tau') \right\}_{\tau' = 1}^{ \tau}$, the goal is to find an optimal policy function $\pi^{\ast}_0$ to take a caching action $\pmb a_0 \left(\tau \!\! +\! 1 \right) = \pi^{\ast}_0 \left(\pmb s_0 \left( \tau \right)\right)$. 

Since requests $\{\pmb r_n (t,\tau)\}_{n,t}$ are Markovian and the present actions $\{\pmb a_n ( t , \tau )\}_{n,t}$ affect future costs, the optimal reinforcement learning policy for the parent node  minimizes the expected discounted cost over all leaf nodes in the long run, namely
\begin{align}\label{eq:pi00}
\pi_0^\ast:=\underset{\pi_0 \in \Pi_0} {\arg\min} \; \mathbb {E} \Big[ \sum_{\tau= 1}^{\infty} \gamma^{\tau-1} {\bf 1}\!^{\top} \! \pmb c_0\! \left( \pmb s_0 (\tau), \pi_0 \! \left( \pmb s_0 (\tau) \right)  \right)  \Big]
\end{align}  
where $\Pi_0$ is set of feasible policies.  


Clearly, the decision at a given state $\pmb s_0 (\tau)$, that is, $\pmb a_0 (\tau +1) = \pi (\pmb s_0(\tau))$, influences the next state $\pmb s_0(\tau+1)$ through $\pi_{n,\tau} (\cdot)$ in \eqref{eq:netstate}, as well as the cost $\pmb c_0 (\cdot)$ in \eqref{eq:c0}. Therefore, problem \eqref{eq:pi00} is a discounted infinite time horizon Markov decision process (MDP). The ensuing section develops a deep reinforcement learning approach to solve this problem.

\subsection{Deep reinforcement learning based caching} \label{sec:RL}

Reinforcement learning deals with action-taking policy function estimation in an environment with dynamically evolving states, so as to minimize a long-term cumulative cost. By interacting with the environment (through successive actions and observed states and costs), reinforcement learning seeks a policy function (of states) to draw actions from, in order to minimize the average cumulative cost as in  \eqref{eq:pi00}~\cite{RLbook}. At this point, it is instructive to recall basic notions of Markov decision processes (MDPs) \cite[p.~310]{RLbook}.

To solve \eqref{eq:pi00}, let us again define the \emph{value function} to indicate the quality of policy $\pi_0$, starting from initial state $\pmb s_0 (0)$ as 
\begin{align}
\label{eq:Vdef}
V_{\pi_0} ( \pmb s_0 ( 0 )):= \mathbb {E} \Big[ \sum_{\tau= 1}^{\infty} \gamma^{\tau-1} {\bf 1}\!^{\top} \pmb \! \pmb c_0 \! \left( \pmb s_0 (\tau), \pi_0 ( \pmb s_0 \left(\tau \right) ) \right) \big| \pmb s_0 ( 0 )  \Big].
\end{align}  

For brevity, the time index and the subscript $0$ referring to the parent node will be dropped whenever it is clear from the context.
To find $V_\pi(\cdot)$, one can rely on the Bellman equation, which basically relates the value of a policy at one state to values of the remaining states \cite[p.~46]{RLbook}. 
Leveraging \eqref{eq:Vdef}, the Bellman equation for value function $V_\pi(\cdot)$ is given by
\begin{align} \label{eq:VBellman} 
V_{\pi} ( \pmb s) =  \mathbb {E} \Big[{\bf 1}\!^{\top} \! {\pmb c} (\pmb s, \pi (\pmb s ))  + \gamma \sum_{\pmb s'} P^{\pi (\pmb s)}_{\pmb s \pmb s'} V_{\pi} (\pmb s') \Big]
\end{align}
where the average immediate cost can be found as 
\begin{align}
\mathbb{E}\! \left[ {\bf 1}\!^{\top} \!{\pmb c} (\pmb s, \pi (\pmb s )) \right] = \sum_{\pmb s'} P^{\pi (\pmb s)}_{\pmb s \pmb s'} {\bf 1}\!^{\top}\! {\pmb c}\! \left(\pmb s, \pi (\pmb s) | \pmb s'\right) 
\end{align}
and $P^{\pi (\pmb s)}_{\pmb s \pmb s'}$ is the underlying transition probability of going from current state $\pmb s$ to the next state $\pmb s'$ upon taking action $\pi(\pmb{s})$. 


Using the value function, we can further define the $Q$-function for policy~$\pi$ as
\begin{align}
\label{eq:Qdef}
Q_{\pi} \! \left(\pmb s, \pmb a\right) := {\mathbb{E}}\! \left[ {\bf 1}\!^{\top}\! {\pmb c} \left(\pmb s, \pmb a \right)  \right] + \gamma \sum_{\pmb s'} P^{\pmb a}_{\pmb s \pmb s'} V_{\pi}\! \left(\pmb s'\right).
\end{align}

As discussed earlier, finding the optimal policy amounts to estimating $Q_{\pi^\ast} (\pmb s, \pmb a)$, which entails estimating a function defined over state and action spaces. In the problem at hand however, the state vector variables are continuous. In this case, (re-)visiting every state-action pair is clearly impossible. To overcome this challenge, we advocate recently popular methods of deep learning to judiciously generalize from only a few observed pairs. Deep learning approaches (see e.g., \cite{sun2019optimization,zhang2019real}) have recently demonstrated remarkable performance in diverse applications, such as object detection, speech recognition, and language translation, to just name a few. DNNs are capable of extracting compact low-dimensional features from high-dimensional data. Wedding DNNs with reinforcement learning, we put forward a deep reinforcement learning approach for the network caching problem, which can effectively deal with the `curse of dimensionality.' 

\begin{figure}
\centering
\includegraphics[width = 0.7 \textwidth]{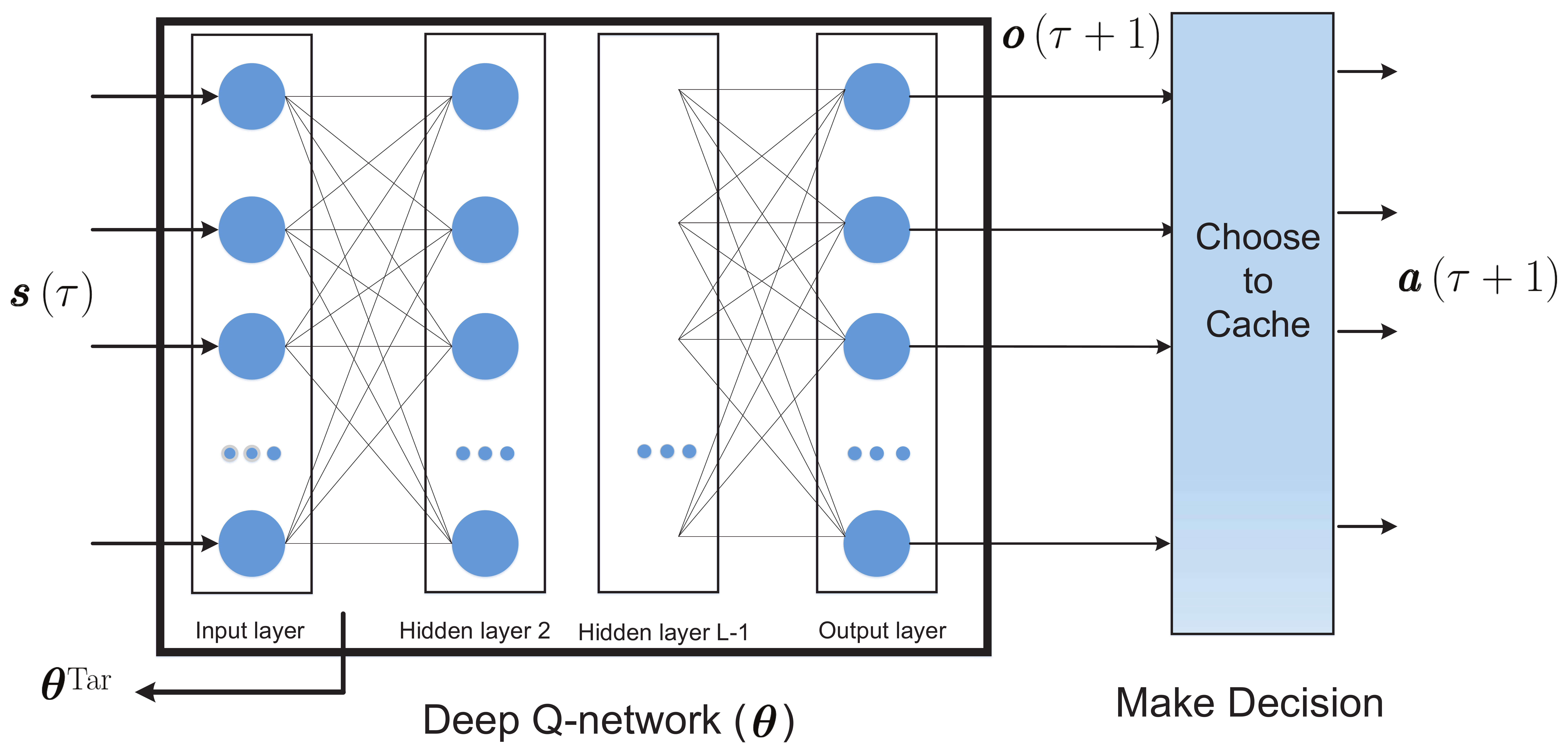}
\caption{Deep $Q$-network}
\label{fig:DQN}
\end{figure}

To this aim, we pursue a parametric model to estimating $Q(\pmb s,  \pmb a)$ with a DNN.  Specifically, we consider a deep feedforward NN having $L$ fully connected layers, with input the $F\times 1$ state vector $\pmb s(\tau)$ as in \eqref{eq:netstate}, and $F\times 1$ output cost vector  $\pmb o (\tau + 1)$~\cite{minh2015}. Each hidden layer $l \in \{ 2, \ldots, L-1\}$ comprises $n_l$ neurons with ReLU activation functions $h(z):=\max(0,\,z)$ for $z \in {\mathbb R}$ (see e.g., \cite{relus} for properties of ReLU networks); Neurons of the $L$-th output layer use a softmax nonlinearity to yield the $f$-th entry of $\pmb o$, an estimated long term cost $o_{f}$, if file $f$ is not stored. Having a limited cache capacity at the parent node $M$ ($<F$), the $M$ largest entries of the DQN output $\pmb o (\tau + 1)$ are to be chosen judiciously by the decision module in Fig.~\ref{fig:DQN} to obtain the action vector $\pmb a (\tau + 1)$. The DNN spits out $\pmb o (\tau + 1)$ as a predicted cost vector for interval $\tau+1$, based on which the caching decision $\pmb a (\tau + 1)$ is found.  

The objective is to train the DQN to find a set of weights collected in the vector~$\boldsymbol \theta$ that parameterizes the input-output relationship $\pmb o (\tau +1 ) = Q(\pmb s (\tau); \boldsymbol \theta_\tau)$. To recursively update 
$\boldsymbol \theta_\tau$ to ${\boldsymbol \theta}_{\tau+1}$, consider two successive intervals along with corresponding states $\pmb s(\tau)$ and $\pmb s(\tau+1)$; the action  $\pmb a(\tau+1)$ taken at the beginning of interval $\tau+1$; and, the cost $\pmb c({\tau}+1)$ revealed at the end of interval $\tau +1$. The instantaneous approximation of the optimal cost-to-go from interval $\tau +1$ is given by $\pmb c({\tau}+1) + \gamma Q(\pmb s(\tau+1) ; \pmb \theta_\tau)$, where $\pmb c({\tau}+1)$ is the immediate cost, 
and $ Q(\pmb s(\tau+1) ; \pmb \theta_\tau)$ represents the predicted cost-to-go from interval $\tau +2$ that is provided by our DQN with $\boldsymbol \theta_\tau$, and discounted by $\gamma$. Since our DQN offers  $Q (\pmb s({\tau}) ; \pmb \theta_{\tau})$ as the predicted cost for interval $\tau+1$, 
the prediction error of this cost as a function of  $\boldsymbol \theta_\tau$ is given by
%
\begin{align}
{\pmb \delta}(\pmb \theta_\tau )  :=  [ \overbrace{\pmb c({\tau}+1) + \gamma Q(\pmb s(\tau+1) ; \pmb \theta_\tau)}^{ \textrm {target cost-to-go from interval $\tau+1$}} - Q (\pmb s({\tau}) ; \pmb \theta_{\tau}) ]  \odot ({\bf 1} - {\pmb a} (\tau +1 ) )
\label{eq:errorDQN}
\end{align}
%
and has non-zero entries for files not stored at interval $\tau+1$. Define for future use the so-termed experience ${\pmb  E}_{\tau+1} := [\pmb s(\tau),   \pmb a(\tau+1), \pmb c(\tau+1), \pmb s(\tau + 1)]$; and adopt as loss function the $\ell_2$-norm of ${\pmb \delta}(\pmb \theta_\tau)$, namely
\begin{align}
\label{eq:lossDQN}
{\mathcal L}  (\pmb \theta_{\tau}) = \big\|{\pmb \delta} (\pmb \theta_\tau ) \big\|_2^2
\end{align}
that DNN weights will be selected to minimize. 
As the cardinality of $\pmb \theta$ is much smaller than $|\mathcal S|  |{\mathcal A}|$, the DQN is effectively trained with few experiences, and generalizes to unseen state vectors. Unfortunately, the DQN model inaccuracy can propagate in the cost prediction error in \eqref{eq:errorDQN} that can in turn cause instability. 
\begin{algorithm}[th!]
\SetKwInOut{Input}{Initialize}
\SetKwInOut{Output}{Output}
\Input{$\pmb s (0)$,  $\pmb s_n (t,\tau), \forall n$, $\pmb \theta_{\tau}$, and $\pmb \theta^{\textrm{Tar}}$ }
\For{$\tau = 1, 2, \ldots $}
{ Take action $\pmb a(\tau)$ via exploration-exploitation 
\\
{ {$\pmb a (\tau) = \bigg\{ \begin{matrix}
{\rm{Best\; files\; via}} \; Q(\pmb s (\tau\!-\!1);\pmb \theta_{\tau}) \hfill {~\rm {w.p.} \; 1- \epsilon_\tau}
\\	{\rm { random } \; \pmb a \in {\mathcal A}}  \hfill {\rm {w.p.} \; \epsilon_\tau} \end{matrix} $} 
\\
\For{$t = 1, \ldots , T$}
{\For{$n = 1, \ldots, N$}{Take action $\pmb a_{n}$ using local policy 
	\\ 
	$\pmb a_n (t, \tau) = \Big\{ \begin{matrix} \!\pi_n (\pmb s ( t -1, \tau )) \; {\rm if} \; t \ne 1
	\\ 
	\pi_n (\pmb s ( T , \tau -1 )) \; {\rm if} \; t = 1 
	\end{matrix} $ 
	\\ 
	{Requests $\pmb r_{n} (t,\tau)$ are revealed 				
		\\ 
		{Set $\pmb s_n (t,\tau) = \pmb r_n (t,\tau)$}
		\\ 
		{Incur $\pmb c_n (\cdot)$}, cf. \eqref{eq:nodalcost}}     	 	
}}
{\bf {Leaf nodes}} 
\\
{Set  \hspace{+.4 cm}${\bar {\pmb s}_n} (\tau):= (1/T) \sum_{t=1}^{T} \; \pmb s_{n} ( t, \tau )$} 
\\ 	
{Send \hspace{+.05 cm} ${\bar {\pmb s}_{n}} ( \tau ) \odot ({\bf 1} - \pi_n ( {\bar {\pmb s}_{n}} ) )$ to parent node}
\\ 
{Send \hspace{+.2 cm}${\bar {\pmb c}}_n (\cdot)$} cf. \eqref{eq:cost_pernode}, to parent node 
\\ 
{\bf {Parent node}} 
\\ 
Set \hspace{+.3 cm} $\pmb s (\tau):= \sum_{n = 1}^{N} w_n  {\bar {\pmb s}_{n}} ( \tau ) \odot ({\bf 1} - \pi_n ( {\bar {\pmb s}_{n}} ) )$ \\ 
Find \,\,   $\pmb c (s({\tau}\!-\!1), \pmb a({\tau}))\;$ 
\\
{{Save \hspace{+.00 cm} $(\pmb s({\tau}\!-\!1), \pmb a({\tau}), \pmb c(s({\tau}\!-\!1), \pmb a({\tau})), \pmb s({\tau}) )$} in $\mathcal E$} 
\\
{Uniformly sample $B$ experiences from $\mathcal E$} 
\\
{Find $  \nabla {\mathcal L}^{\textrm{Tar}} (\boldsymbol \theta) $ for these samples, using \eqref{eq:lossTar} }
\\
{Update~$\boldsymbol \theta_{\tau+1} = \boldsymbol \theta_{\tau} - \beta_\tau \nabla {\mathcal L}^{\textrm{Tar}} (\boldsymbol \theta)$} 
\\
{If $\rm mod (\tau,C) = 0$, then update $\pmb \theta^{\textrm{Tar}} = {\pmb \theta}_\tau$}	
} }
\caption{Deep reinforcement learning for adaptive caching.}
\label{Alg_2}
\end{algorithm}
This is due to: i) correlated experiences used to update the DQN parameters $\pmb \theta$; and, ii) the influence of any changes in policy on subsequent experiences. To handle these challenges, remedies include the so-called {\it experience replay} and {\it target network} to update the DQN weights \cite{minh2015}. In experience replay, the parent node stores all past experiences ${\pmb  E}_{\tau}$ in ${\mathcal E}:= \{\pmb E_{1}, \ldots, \pmb E_{\tau} \}$, and utilizes a batch of $B$ uniformly sampled experiences from this data set, namely $\{\pmb E_{i_\tau}\}_{i= 1}^{B} \sim U (\mathcal E)$. By sampling and replaying previously observed experiences, experience replay overcomes the two challenges. Moreover, to have  target values in \eqref{eq:errorDQN}, a second NN (called target network) with structure identical to the DQN is invoked with parameter vector  $\boldsymbol \theta^{\textrm {Tar}}$. The $\boldsymbol \theta^{\textrm {Tar}}$ will be periodically replaced by $\boldsymbol \theta_{\tau}$ every $C$ training iterations of the DQN. With a randomly sampled experience $\pmb E_{i_\tau}\in\mathcal{E}$, 
the prediction error with the target cost-to-go estimated using the target network is 
\begin{align}
{\pmb \delta}^{\textrm{Tar}}(\pmb \theta ;\pmb E_{i_\tau}) :=   [{\pmb c(i_{\tau}\! +\! 1) + \!  \gamma  Q( {\pmb s(i_{\tau}\! +\! 1)}; {\boldsymbol \theta}^{\textrm {Tar}} )}- Q(\pmb s({i_\tau}) ;\boldsymbol \theta )] \odot ({\bf 1} - \pmb a({i_\tau\! +\! 1}) ). \label{eq:errorTarget}
\end{align}
Different from \eqref{eq:errorDQN}, target values here are found through the target network with weights $\pmb \theta^{\textrm{Tar}}$. In addition, the experience used here is randomly drawn from past experiences in $\mathcal E$. As a result, a reasonable loss function to be optimized is
\begin{align}
\label{eq:lossTar}
{\mathcal L}^{\textrm{Tar}} (\pmb \theta) = {\mathbb E} \| {\pmb \delta}^{\textrm{Tar}}(\pmb \theta; \pmb E ) \|_2^2
\end{align}
where the expectation is taken with respect to the uniformly sampled experience $\pmb E$. In practice however, only a batch of $B$ samples is available to update $\pmb \theta_{\tau}$, so the expectation will be replaced by the sample average. Finally, upon minimizing the loss in \eqref{eq:lossTar} the DQN weights will be updated. Incorporating these remedies, Algorithm \ref{Alg_2} tabulates our deep reinforcement learning based adaptive caching scheme. 
	\begin{figure}[t!]
		\centering
		\includegraphics[width=.5 \textwidth]{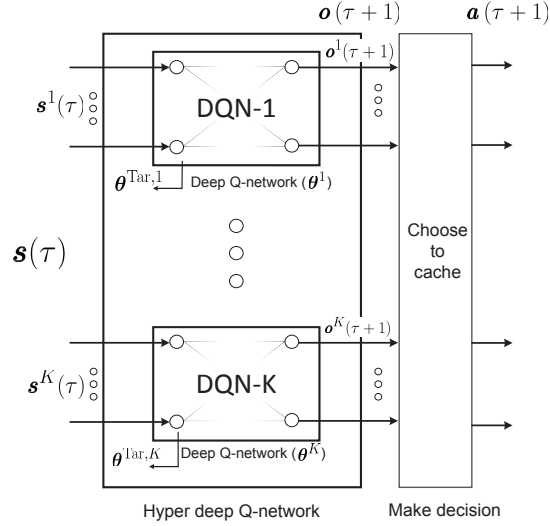}\caption{Hyper deep-Q network for scalable caching.}
		\label{fig:H_DQN}
	\end{figure}

\subsection{Numerical tests} 
\label{sec:numerical}
Here, we present several numerical tests to assess the performance of the proposed deep reinforcement learning based caching scheme. We consider a parent node with $N = 5$ leaf nodes, where every leaf node implements a local caching policy $\pi_n$ with capacity to store $M_n = 5$ files. Every leaf node receives end-user requests per fast time slot, and each slow-timescale interval contains $T = 2$ slots. Content popularity exhibits different Markovian dynamics locally at leaf nodes. Having no access to local policies $\{\pi_n\}$, 
the parent node not only should learn file popularities along with their temporal evolutions, but also learn the caching policies of leaf nodes. To endow our approach with scalability to handle $F \gg 1$, we advocate hyper $Q$-network implementation, where files are first split into $K$ smaller groups of sizes $F_1,  \ldots, F_K$ with $F_k\ll F$. This yields the representation $\pmb s^\top(\tau) := [{\pmb s^{1}}^\top\!(\tau ) , \ldots , {\pmb s^{K}}^\top\!(\tau ) ]$, where $\pmb s^k \in {\mathbb R}^{F_k}$. By running $K$ DQNs in parallel, every DQN-$k$ now spits out the associated predicted costs of input files through $\pmb o^k (\tau) \in {\mathbb R}^{F^k}$. Concatenating all these outputs, one obtains the predicted output cost vector of all files as $\pmb o^\top(\tau+1)\!:=\! [  {{\pmb o}^1}^\top\!(\tau+1), \ldots , {{\pmb o}^K}^\top\!(\tau+1)]$; see Fig. \ref{fig:H_DQN}.  
\begin{figure}[t!]
	\centering
	{\includegraphics[width=0.485\textwidth]{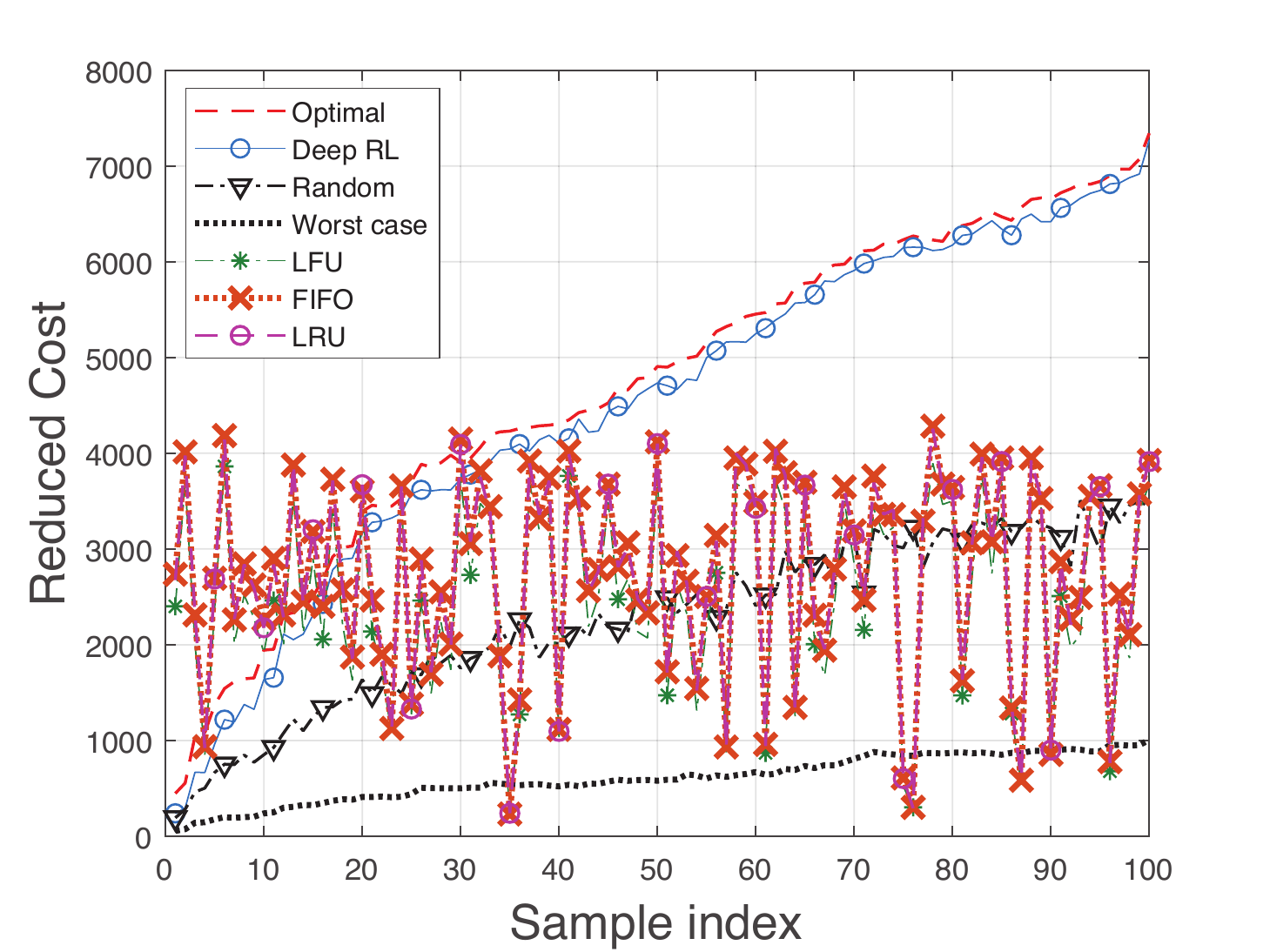}}
	\hspace{.02 cm}
	{\includegraphics[width=0.485\textwidth]{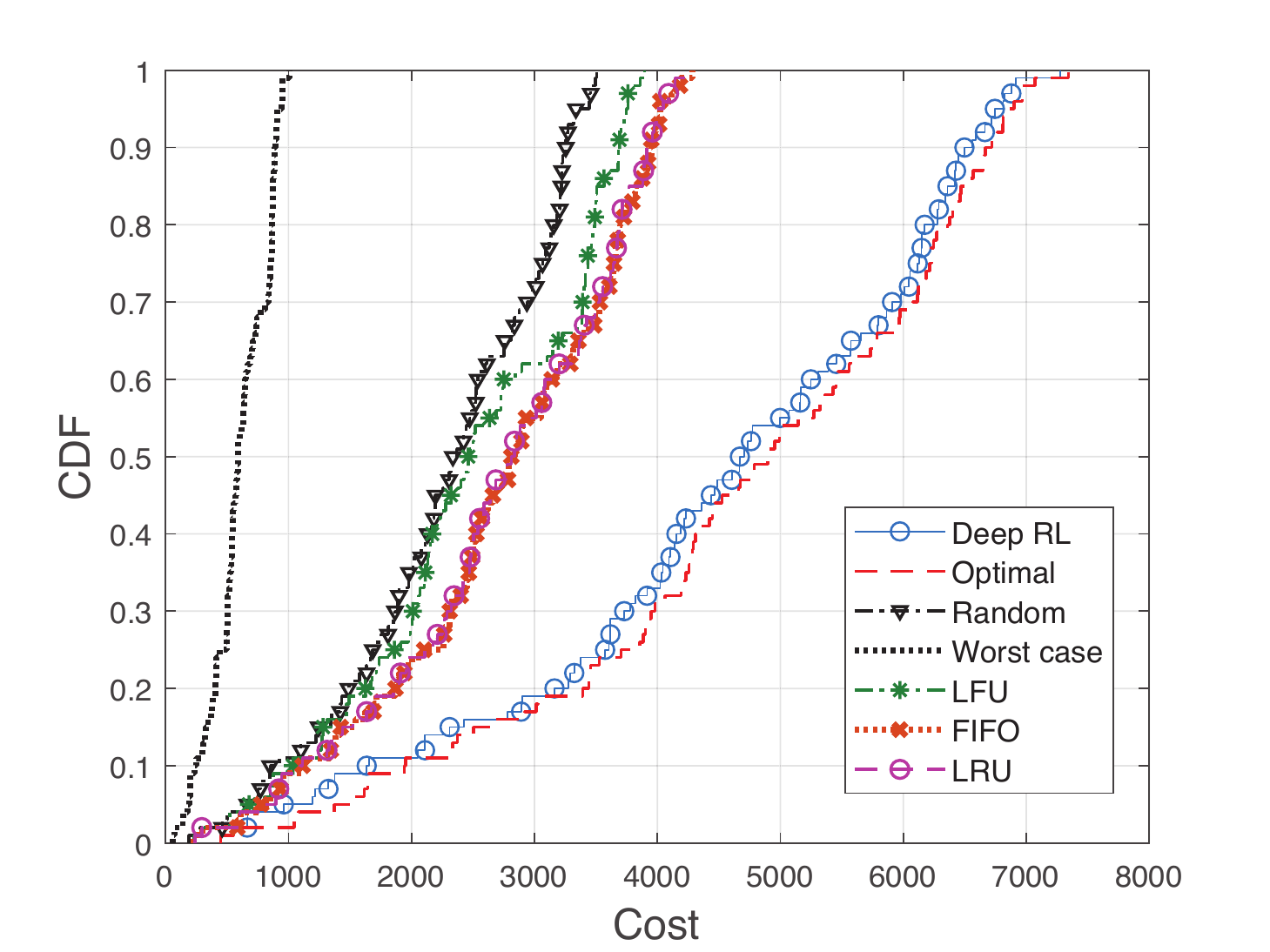} }
	\caption{Instantaneous reduced cost for different policies (left);
		CDF of reduced cost for different policies (right).}
	\label{fig:scalable_2}
\end{figure}

\begin{figure}[t!]
	\centering
	{\includegraphics[width=0.485\textwidth]{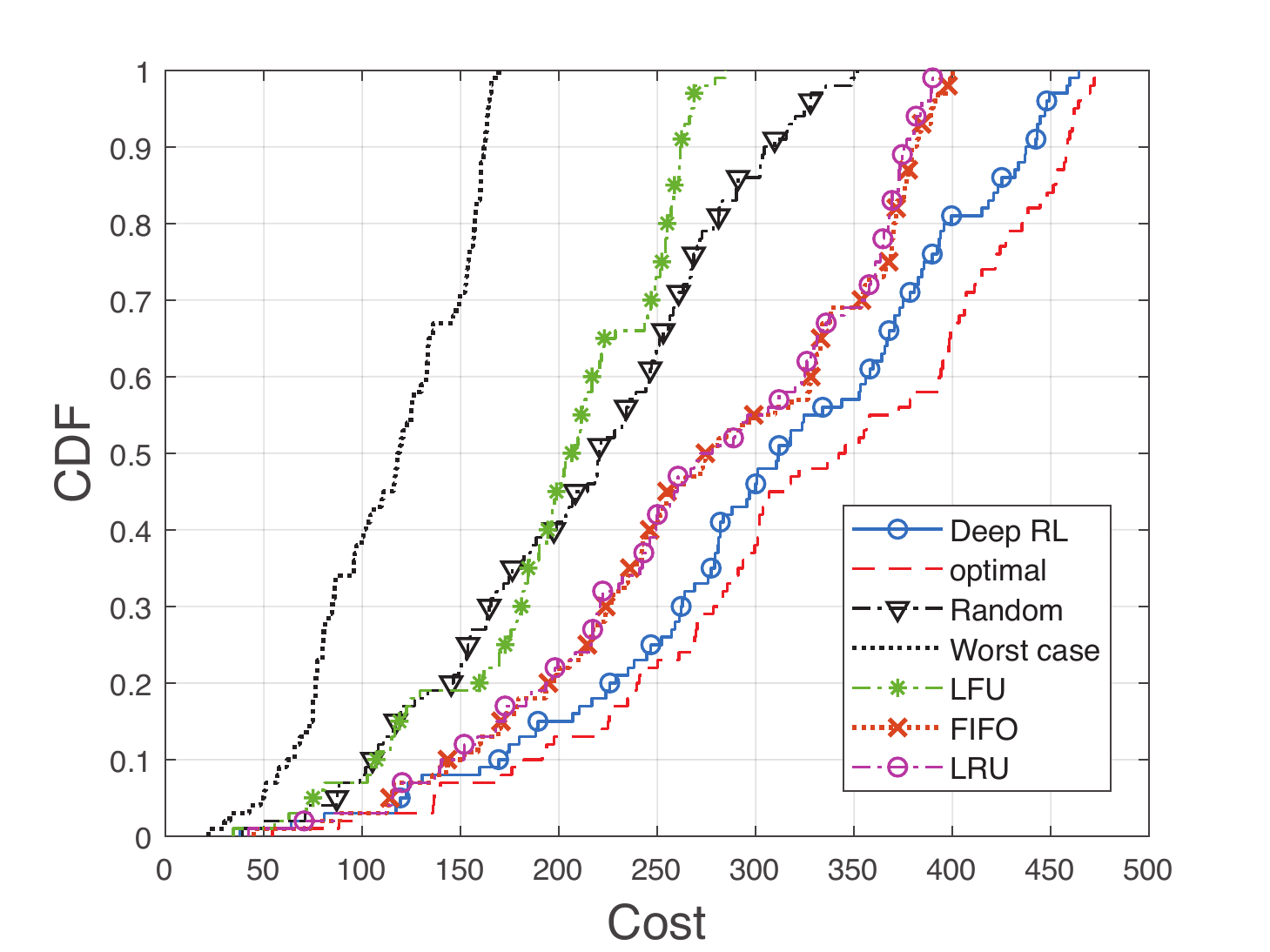}}
	\hspace{.02 cm}
	{\includegraphics[width=0.485\textwidth]{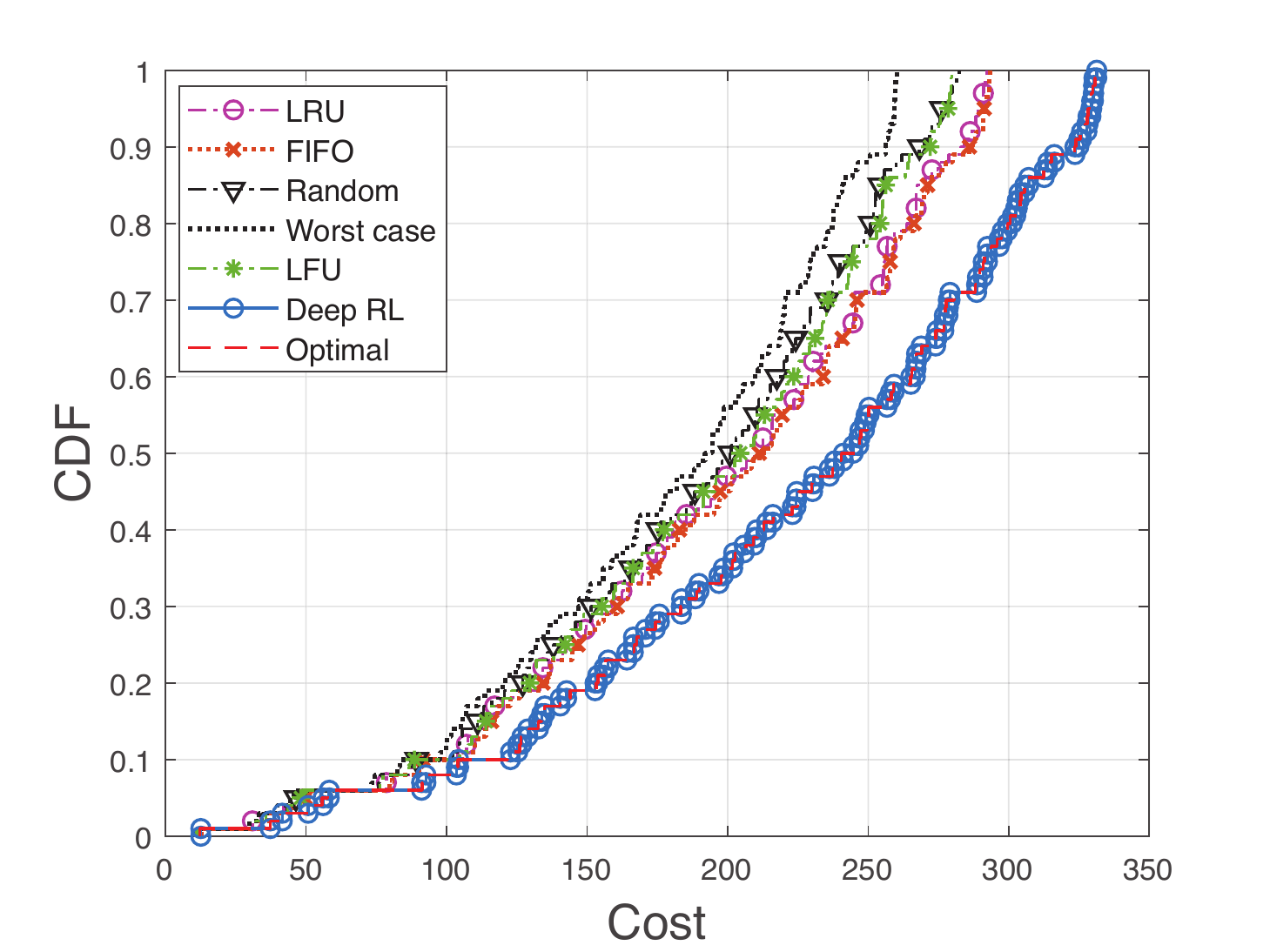} }
	\caption{Performance for $N = 10$ nodes (left), and $N = 1,000$ leaf nodes (right).}
	\label{fig:N_nodes}
\end{figure}

The numerical tests consider $F = 1,000$ files with $K = 25$ and $\{F_k = 20\}$, where only $M_0 = 50$ can be stored. To further assess the performance, we adopt a \textit{non-causal} optimal policy as a benchmark, which assumes having knowledge of future requests and stores the most frequently requested files. In fact, this is the best policy that one may ideally implement. Further, practically used cache  schemes including the LRU, LFU, and FIFO~\cite{dan1990approximate} are also simulated. A difference between LRU, LFU, FIFO, and our proposed approach is that in these three schemes, the cache is refreshed whenever a request is received, while in our scheme it is only refreshed at the end of every time slot. By drawing $100$ samples randomly from all $2,000$ time intervals, the instantaneous reduced cost and the empirical cumulative distribution function (CDF) obtained over these $100$ random samples are depicted in Fig.~\ref{fig:scalable_2}. These plots further illustrate how the deep reinforcement learning policy performs relative to the alternatives, and in particular it approximates closely the optimal policy. LRU, LFU, and FIFO make caching decisions based on instantaneous observations, and can refresh the cache many times within each slot. Yet, our proposed policy as well as the optimal one here learn from all historical observations to cache, and refresh the cache only once per slot. Because of this difference, the former policies outperform the latter at the very beginning of Fig. \ref{fig:scalable_2} (left), but they do not adapt to the underlying popularity evolution and are outperformed by our learning-based approach after a number of slots. The merit of our approach is further illustrated by the CDF of the reduced cost depicted in Fig.~\ref{fig:scalable_2} (right).

In the last test, the number of leaf nodes is increased from $10$ in the previous experiment to~$N = 1,000$. Figures~\ref{fig:N_nodes}(left) and \ref{fig:N_nodes}(right) showcase that the deep reinforcement learning performance approaches that of the optimal one as the number of nodes increases. This is likely because the more leaf nodes there are, the smoother the popularity fluctuations become, and the easier it is for deep reinforcement learning to learn the optimal policy.

\section{Discussion}
This chapter has reviewed some recent reinforcement learning-enabled caching approaches for both single- and multi-node settings. After carefully accounting for space-time popularities, conventional reinforcement learning based policies were introduced. To overcome the challenges arising due to the high dimensionality in the state- or action- spaces, function approximation schemes were developed. In particular, linear function approximation as well as deep neural network based models were advocated. Although numerical tests highlighted the merits of the novel approaches, several practical challenges remain in our future research agenda. For instance, bringing computation and communication resources closer to the end users along with caching will further enhance network performance. However, jointly optimizing available resources is a formidable challenge. In this context, innovative resource allocation schemes are needed, where toolkits such as e.g., graph-based representations, non-convex optimization schemes, state-of-the-art machine learning methods, statistical signal processing techniques, sparse and low-rank models only to name a few, can play critical roles to realize the vision of smart networking.     


\noindent 
{\bf Acknowledgement}. The work in this chapter was supported by the NSF grants 1508993, 1711471, and 1901134.
\bibliographystyle{IEEEtran}
\bibliography{sample-vancouver}

\end{document}